\RequirePackage{rotating}
\documentclass[useAMS,usenatbib]{mnras}

\usepackage{ulem}
\usepackage{color}
\usepackage{graphics,graphicx}
\usepackage{times} 
\usepackage{amssymb}
\usepackage{amsmath}
\usepackage{natbib}
\usepackage{lscape}
\usepackage{url}
\usepackage{multirow}
\usepackage{ctable}
\usepackage{threeparttable}
\usepackage[labelfont=bf,labelsep=period]{caption}
\newif\ifAMStwofonts
\AMStwofontstrue

\def\xmm{{\it XMM-Newton}}

\def\chandra{{\it Chandra}}

\def\heg{HEG}
\def\hetg{HETG}
\def\hetgs{HETGS}
\def\meg{MEG}
\def\swift{{\it Swift}}
\def\swiftng{{\it Neil Gehrels Swift Observatory}}

\def\epicpn{{EPIC-pn}}
\def\epicmos1{{EPIC-MOS1}}
\def\epicmos2{{EPIC-MOS2}}
\def\epicmos{{EPIC-MOS}}

\def\nustar{{\it NuSTAR}}

\def\deg{$^{\circ}$}

\def\pcmsq{\hbox{$\rm\thinspace cm^{-2}$}}
\def\H0{{\rm ~km~s^{-1}~Mpc^{-1}}}

\def\ergpcmsqps{\hbox{$\rm\thinspace erg~cm^{-2}~s^{-1}$}}

\def\ergps{\hbox{erg~s$^{-1}$}}

\def\msun{\hbox{$\rm M_{\odot}$}}

\def\mdot{$\dot{m}$}

\def\chisq{{$\chi^{2}$}}

\def\xspec{\hbox{\small XSPEC}}
\def\isis{\hbox{\small ISIS}}

\def\ciao{\hbox{\rm{\small CIAO}}}

\def\heasoft{\hbox{\rm{\small HEASOFT}}}
\def\nustardas{\rm {\small NUSTARDAS}}

\def\xmmselect{\hbox{\rm{\small XMMSELECT}}}
\def\ftool{\hbox{\rm{\small FTOOL}}}

\def\specgroup{\rm {\small SPECGROUP}}

\def\addascaspec{\hbox{\rm{\small ADDASCASPEC~\/}}}
\def\flx2xsp{\rm{\small FLX2XSP}}
\def\sas{\hbox{\rm{\small SAS~\/}}}
\def\epchain{\hbox{\rm{\small EPCHAIN}}}
\def\emchain{\hbox{\rm{\small EMCHAIN}}}

\def\rmfgen{\hbox{\rm{\small RMFGEN}}}
\def\arfgen{\hbox{\rm{\small ARFGEN}}}

\def\addascaspec{\rm{\small ADDASCASPEC}}
\def\nupipeline{\rm{\small NUPIPELINE}}
\def\nuproducts{\rm{\small NUPRODUCTS}}

\def\stingray{\hbox{\rm {\small STINGRAY}}}
\def\hendrics{\hbox{\rm {\small HENDRICS}}}
\def\presto{\hbox{\rm {\small PRESTO}}}

\def\grid25{\hbox{\rm{\small GRID25}}}

\def\simpl{\rm{\small SIMPL}}

\def\cabs{\rm{\small CABS}}
\def\tbabs{\rm{\small TBABS}}

\def\diskbb{\rm{\small DISKBB}}
\def\diskpbb{\rm{\small DISKPBB}}

\def\cutoffpl{\rm{\small CUTOFFPL}}

\def\eg{{\it e.g.}}

\def\ie{{\it i.e.~\/}}

\def\la{\mathrel{\hbox{\rlap{\hbox{\lower4pt\hbox{$\sim$}}}{\raise2pt\hbox{$<$}}}}}
\def\ga{\mathrel{\hbox{\rlap{\hbox{\lower4pt\hbox{$\sim$}}}{\raise2pt\hbox{$>$}}}}}

\def\d25{D$_{25}$}

\def\.25{0.25 keV\thinspace}

\def\rg{$R_{\rm{G}}$}

\def\risco{$R_{\rm{ISCO}}$}
\def\rin{$R_{\rm in}$}

\def\rsp{$R_{\rm{sp}}$}
\def\rtrap{$R_{\rm{trap}}$}
\def\rmag{$R_{\rm{M}}$}

\def\rns{$R_{\rm{NS}}$}

\def\ngc{NGC\,1313 X-1}

\title[The Unusual Spectral Variability of NGC\,1313 X-1]{The Unusual
Broadband X-ray Spectral Variability of NGC\,1313 X-1 seen with \textit{XMM-Newton},
\textit{Chandra} and \textit{NuSTAR}}

\author[D.\,J. Walton et al.]
{\parbox{7.in}{D.\,J. Walton$^{1}$\thanks{E-mail: dwalton@ast.cam.ac.uk},
C. Pinto$^{2,3}$, % DONE, DONE
M. Nowak$^{4}$, % DONE
M. Bachetti$^{5}$,
R. Sathyaprakash$^{6}$, % DONE
E. Kara$^{7}$, % DONE
\\[0.05cm]
T. P. Roberts$^{6}$, % DONE
R. Soria$^{8,9}$, % DONE
M. Brightman$^{10}$, % DONE
C. R. Canizares$^{7}$, % DONE
H. P. Earnshaw$^{10}$, % DONE
\\[0.05cm]
F. F\"urst$^{11}$, % DONE
M. Heida$^{12}$, % DONE
M. J. Middleton$^{13}$,
D. Stern$^{14}$,
L. Tao$^{15}$, % DONE
N. Webb$^{16}$, % DONE
\\[0.05cm]
W. N. Alston$^{1}$,
D. Barret$^{16}$,
A. C. Fabian$^{1}$,
F. A. Harrison$^{10}$,
P. Kosec$^{1}$
\\[0.3cm]
\footnotesize
$^{1}$ \it{Institute of Astronomy, University of Cambridge, Madingley Road, Cambridge CB3 0HA, UK} \\
$^{2}$ \it{ESTEC/ESA, Keplerlaan 1, 2201AZ Noordwijk, The Netherlands} \\ 
$^{3}$ \it{INAF - IASF Palermo, Via U. La Malfa 153, I-90146 Palermo, Italy} \\
$^{4}$ \it{Physics Department, CB 1105, Washington University, One Brookings Drive, St.  Louis, MO 63130-4899} \\
$^{5}$ \it{INAF-Osservatorio Astronomico di Cagliari, via della Scienza 5, I-09047 Selargius, Italy} \\
$^{6}$ \it{Centre for Extragalactic Astronomy, Durham University, Department of Physics, South Road, Durham DH1 3LE, UK} \\
$^{7}$ \it{MIT Kavli Institute for Astrophysics and Space Research, Cambridge, MA 02139, USA} \\ 
$^{8}$ \it{College of Astronomy and Space Sciences, University of the Chinese Academy of Sciences, Beijing 100049, China} \\
$^{9}$ \it{Sydney Institute for Astronomy, School of Physics A28, The University of Sydney, Sydney, NSW 2006, Australia} \\ 
$^{10}$ \it{Space Radiation Laboratory, California Institute of Technology, Pasadena, CA 91125, USA} \\
$^{11}$ \it{European Space Astronomy Centre (ESA/ESAC), Operations Department, Villanueva de la Ca\~nada (Madrid), Spain} \\
$^{12}$ \it{European Southern Observatory, Karl-Schwarzschild-Stra$\beta$e 2, 85748 Garching bei M\"unchen, Germany} \\
$^{13}$ \it{Department of Physics and Astronomy, University of Southampton, Highfield, Southampton SO17 1BJ, UK} \\
$^{14}$ \it{Jet Propulsion Laboratory, California Institute of Technology, Pasadena, CA 91109, USA} \\
$^{15}$ \it{Key Laboratory of Particle Astrophysics, Institute of High Energy Physics, Chinese Academy of Sciences, Beijing 100049, China} \\
$^{16}$ \it{IRAP, Universit\'e de Toulouse, CNRS, CNES, 9 avenue du Colonel Roche, Toulouse, France} \\
}}
\date{}

\begin{document}
\pagerange{\pageref{firstpage}--\pageref{lastpage}}
\maketitle
\label{firstpage}

%\vspace*{-0.5cm}
\begin{abstract}
We present results from the major coordinated X-ray observing program on the ULX
\ngc\ performed in 2017, combining \xmm, \chandra\ and \nustar, focusing on the
evolution of the broadband ($\sim$0.3--30.0\,keV) continuum emission. Clear and
unusual spectral variability is observed, but this is markedly suppressed above
$\sim$10--15\,keV, qualitatively similar to the ULX Holmberg\,IX X-1. We model the
multi-epoch data with two-component accretion disc models designed to approximate
super-Eddington accretion, allowing for both a black hole and a neutron star accretor.
With regards to the hotter disc component, the data trace out two distinct tracks in the
luminosity--temperature plane, with larger emitting radii and lower temperatures seen
at higher observed fluxes. Despite this apparent anti-correlation, each of these tracks
individually shows a positive luminosity--temperature relation. Both are broadly
consistent with $L\propto{T}^{4}$, as expected for blackbody emission with a constant
area, and also with $L\propto{T}^{2}$, as may be expected for an advection-dominated
disc around a black hole. We consider a variety of possibilities for this unusual
behaviour. Scenarios in which the innermost flow is suddenly blocked from view by
outer regions of the super-Eddington disc/wind can explain the luminosity--temperature
behaviour, but are difficult to reconcile with the lack of strong variability at higher
energies, assuming this emission arises from the most compact regions. Instead, we
may be seeing evidence for further radial stratification of the accretion flow than is
included in the simple models considered, with a combination of winds and advection
resulting in the suppressed high-energy variability.
\end{abstract}


\section{Introduction}

\ngc\ is one of the archetypal ultraluminous X-ray sources (ULXs). These are
off-nuclear X-ray sources that appear to radiate in excess of $10^{39}$\,\ergps,
roughly the Eddington limit for the stellar remnant black holes ($M_{\rm{BH}} \sim
10$\,\msun) that power X-ray binaries in our own Galaxy (see \citealt{Kaaret17rev}
for a recent review). Although they were historically considered good candidates for
hosting intermediate mass black holes (IMBHs, $M_{\rm{BH}} \sim 10^{2-5}$\,\msun;
\eg\ \citealt{Miller03, Miller04hoIX, MilCol04, Strohmayer09a}), the broadband X-ray
spectra observed in the \nustar\ era show clear deviations from the standard
sub-Eddington accretion modes which would be expected for such objects
(\eg\ \citealt{Bachetti13, Walton13culx, Walton14hoIX, Walton15, Walton15HoII, 
Walton17hoIX, Sazonov14, Rana15, Mukherjee15, Annuar15, Luangtip16,
Krivonos16, Fuerst17ngc5907, Shidatsu17}). These observations confirmed the
previous indications for these deviations seen in the more limited \xmm\ bandpass
(\citealt{Stobbart06, Gladstone09}), and imply instead that the majority of ULXs likely
represent a population of X-ray binaries accreting at high/super-Eddington rates.

This conclusion was further cemented with the remarkable discovery that a number of
ULXs are actually powered by accreting pulsars (\citealt{Bachetti14nat, Fuerst16p13,
Israel17, Israel17p13, Carpano18, Sathyaprakash19, Rodriguez19}). These sources
therefore appear to exceed their Eddington limits by factors of up to $\sim$500, and
their broadband spectra are qualitatively similar to the rest of the ULX population,
particularly at high energies ($E > 10$\,keV; \citealt{Pintore17, Koliopanos17,
Walton18p13, Walton18ulxBB}). Two other ULXs in M51 have also been identified as
likely hosting neutron star accretors via other means, first through the detection of a
potential cyclotron resonant scattering feature (\citealt{Brightman18}), and second
through the detection of a potentially bi-modal flux distribution (as expected for a
source transitioning to and from the propeller regime; \citealt{Earnshaw18}). A number
of other candidates for such transitions have also recently been reported among the
broader ULX population (\citealt{Song20}). Although the known ULX pulsars are
now firmly established as being powered by magnetised neutron stars, the strength of
the magnetic fields in these systems is still the subject of significant debate, with
estimates ranging from $\sim$10$^{9-15}$\,G and some models invoking higher-order
field geometries (\eg\ quadrupolar) than the standard dipole fields typically assumed
(\eg\ \citealt{Bachetti14nat, Dallosso15, Mushtukov15, Kluzniak15, Fuerst16p13,
Brightman18, Walton18crsf, Vasilopoulos18, Middleton19}).

One of the fundamental predictions of super-Eddington accretion is that strong winds
are launched from the accretion flow, as radiation pressure exceeds gravity
(\eg\ \citealt{Shakura73, Poutanen07, Dotan11, Takeuchi13}). Observational evidence
for such winds in ULXs was only recently seen for the first time in \ngc, through the
detection of strongly blueshifted atomic features (\citealt{Pinto16nat, Walton16ufo};
see also \citealt{Middleton15soft}), and such features have now been seen in several
other systems (including NGC\,300 ULX1, one of the known pulsars; \citealt{Pinto17, 
Kosec18ulx, Kosec18}). These winds are extreme, reaching velocities of up to
$\sim$0.25$c$, and, despite the extreme X-ray luminosities of these sources, may
actually dominate their total energetic output.

In order to study the variability of the extreme outflow in \ngc, we performed a series
of coordinated broadband X-ray observations of the NGC\,1313 galaxy combining data
from \xmm\ (PI: Pinto), \chandra\ (PI: Canizares) and \nustar\ (PI: Walton). This
program has already provided a variety of interesting results, revealing clear variability
in the wind in X-1 (\citealt{Pinto20}), a $\sim$150\,s soft X-ray lag in X-1
(\citealt{Kara20}), and the detection of pulsations in X-2 (\citealt{Sathyaprakash19}).
Here we combine these observations with the coordinated \xmm\ and \nustar\
observations available in the archive to study the broadband continuum variability
exhibited by \ngc.

\begin{table}
  \caption{Details of the X-ray observations of NGC\,1313 used in this work.}
\begin{center}
\vspace{-0.4cm}
\begin{tabular}{c c c c c}
\hline
\hline
\\[-0.2cm]
Epoch & Mission(s) & OBSID(s) & Start & Exposure\tmark[a] \\
\\[-0.3cm]
& & & Date & (ks) \\
\\[-0.3cm]
\hline
\hline
\\[-0.2cm]
\multirow{4}{*}{XN1} & \nustar\ & 30002035002 & 2012-12-16 & 154 \\
& \nustar\ & 30002035004 & 2012-12-21 & 206 \\
& \xmm\ & 0693850501 & 2012-12-16 & 85/112 \\
& \xmm\ & 0693851201 & 2012-12-22 & 79/120 \\
\\[-0.1cm]
\multirow{2}{*}{XN2} & \nustar\ & 80001032002 & 2014-07-05 & 73 \\
& \xmm\ & 0742590301 & 2014-07-05 & 54/61 \\
\\[-0.1cm]
\multirow{2}{*}{XN3} & \nustar\ & 90201050002 & 2017-03-29 & 125 \\
& \xmm\ & 0794580601 & 2017-03-29 & 25/38 \\
\\[-0.1cm]
\multirow{3}{*}{XN4} & \nustar\ & 30302016002 & 2017-06-14 & 100 \\
& \xmm\ & 0803990101 & 2017-06-14 & 110/130 \\
& \xmm\ & 0803990201 & 2017-06-20 & 111/129 \\
\\[-0.1cm]
\multirow{8}{*}{CN1} & \nustar\ & 30302016004 & 2017-07-17 & 87 \\
& \chandra\ & 19929 & 2017-07-03 & 18 \\
& \chandra\ & 20105 & 2017-07-06 & 31 \\
& \chandra\ & 19712 & 2017-07-18 & 49 \\
& \chandra\ & 20125 & 2017-08-01 & 25 \\
& \chandra\ & 20126 & 2017-08-02 & 27 \\
& \chandra\ & 19714 & 2017-08-03 & 20 \\
& \chandra\ & 19927 & 2017-08-05 & 25 \\
\\[-0.1cm]
\multirow{3}{*}{XN5} & \nustar\ & 30302016006 & 2017-09-03 & 90 \\
& \xmm\ & 0803990301 & 2017-08-31 & 45/101 \\
& \xmm\ & 0803990401 & 2017-09-02 & 83/81 \\
\\[-0.1cm]
\multirow{6}{*}{CN2} & \nustar\ & 30302016008 & 2017-09-15 & 108 \\
& \chandra\ & 19928 & 2017-08-28 & 21 \\
& \chandra\ & 20729 & 2017-09-16 & 10 \\
& \chandra\ & 20637 & 2017-09-24 & 28 \\
& \chandra\ & 19713 & 2017-09-26 & 32 \\
& \chandra\ & 20798 & 2017-09-29 & 17 \\
\\[-0.1cm]
\multirow{2}{*}{XN6\tmark[b]} & \nustar\ & 30302016010 & 2017-12-09 & 100 \\
& \xmm\ & 0803990601 & 2017-12-09 & 69/115 \\
\\[-0.3cm]
\hline
\hline
\\[-0.4cm]
\end{tabular}
\end{center}
$^{a}$ Good exposures used for our spectral analysis, given to the nearest ks; \xmm\
exposures are listed for the \epicpn/MOS detectors, and the \nustar\ exposures
combine both modes 1 and 6 (see Section \ref{sec_nustar}). \\
$^{b}$ Although two \xmm\ observations were taken as part of epoch XN6, the two
show markedly different spectra, so we only make use of the \xmm\ observation taken
simultaneously with the \nustar\ exposure for this epoch (see Section \ref{sec_epochs}).
\label{tab_obs}
\end{table}

\begin{figure*}
\begin{center}
\hspace*{-0.3cm}
\rotatebox{0}{
{\includegraphics[width=500pt]{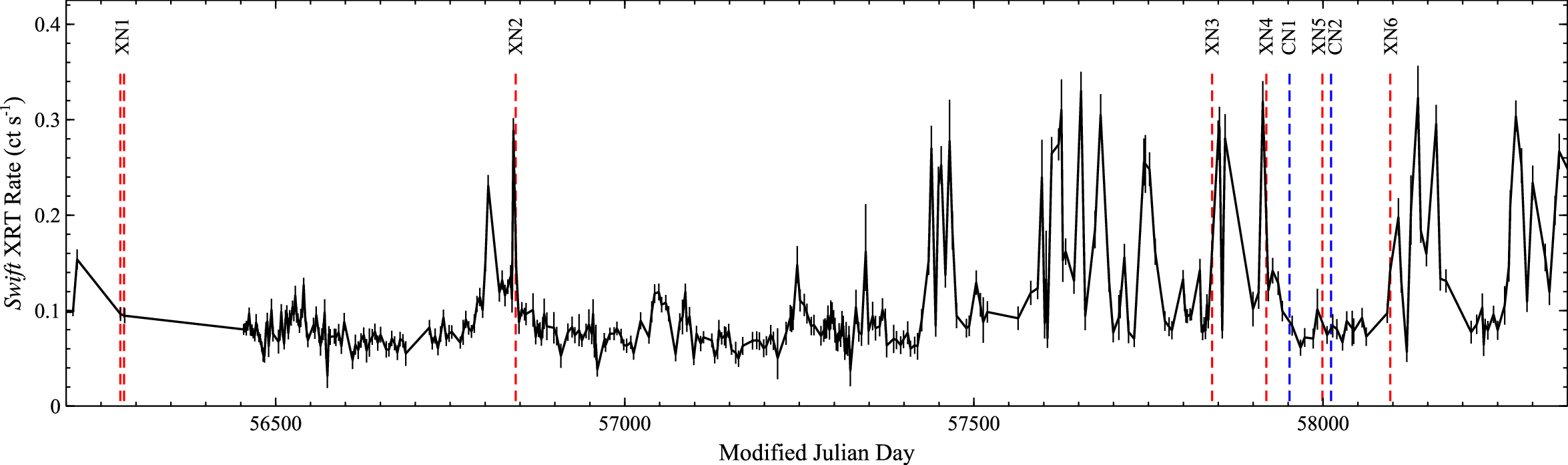}}
}
\end{center}
\vspace*{-0.3cm}
\caption{
The long-term 0.3--10\,keV X-ray lightcurve seen by \swift\ XRT since October 2012. The
timing of the \nustar\ observations, which define the broad epochs considered here, are
indicated by the red dashed lines for coordination with \xmm, and blue dashed lines for
coordination with \chandra. The period after MJD $\sim$ 57400 shows enhanced
variability, which is covered by a number of our broadband observations.}
\label{fig_longlc}
\end{figure*}

\section{Observations and Data Reduction}
\label{sec_red}

During the main 2017 campaign, \xmm\ (\citealt{XMM}) observed NGC\,1313 with its
full suite of instrumentation for six full orbits ($\sim$750\,ks total exposure), grouped
into three pairs of observations, with the two observations constituting each pair taken
in relatively quick succession. \chandra\ (\citealt{CHANDRA}) also performed a series
of observations across 2017, totalling $\sim$500\,ks exposure. Although these
\chandra\ observations are relatively well spread across the year, there are two periods
where a number of the observations are clustered in time. In addition to these soft
X-ray observations, \nustar\ (\citealt{NuSTAR}) also performed a series of five
exposures over the second half of 2017 ($\sim$500\,ks total exposure). One was
performed with each of the pairs of \xmm\ observations, taken simultaneously with one
of the two exposures, and the remaining two were taken with the two main clusters of
\chandra\ observations, again performed simultaneously with one of the exposures in
these two groups.

In addition to these observations, \xmm\ and \nustar\ have also performed coordinated
observations on a few prior occasions. There were two deep observations taken in
quick succession soon after the launch of \nustar\ in 2012 (see \citealt{Bachetti13,
Miller14, Walton16ufo}), as well as shorter coordinated observations taken in 2014 and
again in 2017 prior to the commencement of the main campaign. Details of the
observations considered in this work are given in Table \ref{tab_obs}, and the following
sections describe our data reduction procedure for the various missions involved.

\subsection{\textit{NuSTAR}}
\label{sec_nustar}

We reduced the \nustar\ data as standard with the \nustar\ Data Analysis Software
(v1.8.0; part of the \heasoft\ distribution) and \nustar\ caldb v20171204. The unfiltered
event files were cleaned with \nupipeline. We used the standard depth correction,
which significantly reduces the internal background at high energies, and periods of
earth-occultation and passages through the South Atlantic Anomaly were also
excluded. To be conservative, source products were extracted from circular regions of
radius 30$''$, owing to the presence of another faint source to the south (see
\citealt{Bachetti13}), and the background was estimated from a larger, blank area on
the same detector, free of contaminating point sources (the exact size of the background
region used varies between epochs, depending on the position angle of the observation
and the proximity of the source to the edge of the detector). Spectra and lightcurves were
extracted from the cleaned event files using \nuproducts\ for both focal plane modules
(FPMA and FPMB). In addition to the standard `science' data (mode 1), we also extract
the `spacecraft science' data (mode 6) following \cite{Walton16cyg}. This provides
$\sim$20--40\% of the total good exposure, depending on the specific observation.

\subsection{\textit{XMM-Newton}}

For the \xmm\ observations, data reduction was carried out with the \xmm\ Science
Analysis System (\sas v16.1.0). Here we focus on the data taken by the EPIC CCD
detectors -- \epicpn\ and \epicmos\ (\citealt{XMM_PN, XMM_MOS}) -- for our
continuum analysis; the high-resolution data from the RGS (\citealt{XMM_RGS}) will
be presented in a separate work (\citealt{Pinto20}). The raw observation data files were
processed as standard using \epchain\ and \emchain. Source products were extracted
from circular regions of radius $\sim$30$''$, and, as before, the background was
estimated from larger areas on the same CCD free from contaminating point sources
(again, the exact size of the background region used varies from epoch to epoch,
depending on the position angle of the observation and, for the \epicpn\ detector, the
position of the read-out streak relative to other nearby sources). Lightcurves and
spectra were generated with \xmmselect, excluding periods of high background and
selecting only single and double events for \epicpn\ ({\small PATTERN}$\leq$4), and
single to quadruple events for \epicmos\ ({\small PATTERN}$\leq$12). The
redistribution matrices and auxiliary response files for each detector were generated
with \rmfgen\ and \arfgen, respectively. After performing the data reduction separately
for each of the MOS detectors, and confirming their consistency, these spectra were
combined using the \ftool\ \addascaspec\ for each OBSID. Finally, where spectra from
different OBSIDs were merged, these were also combined using \addascaspec.

\subsection{\textit{Chandra}}

High spectral resolution X-ray spectroscopy is provided by the \chandra-High Energy
Gratings Spectrometer \citep[\hetgs][]{CHANDRA_HETG}, which consists of two
gratings sets: the Medium Energy Gratings (\meg, $E/\Delta E \approx 700$ at 1\,keV)
and the High Energy Gratings (\heg, $E/\Delta E \approx 1300$ at 1\,keV). Each
gratings set, \meg\ and \heg, disperses spectra into negative and positive orders.
Throughout this work, as we describe below, we combine the $\pm1^{\rm st}$ order
spectra into a single spectrum for each gratings instrument (appreciable counts in the
higher orders are only obtained for very bright sources, so are ignored in our analyses). 
The \hetg\ also produces an undispersed $\mathrm{0^{th}}$ order spectrum, but it
suffers from photon pileup \citep{PILEUP}, so we do not consider it further.

The \chandra-\hetg\ spectra were created using the Interactive Spectral Interpretation
System (\isis; \citealt{ISIS}) running extraction scripts from the Transmission
Gratings Catalogue \citep[TGcat;][]{TGcat}. These scripts in turn ran tools from \ciao\
v4.10 and utilized \chandra\ caldb v4.7.8. The position of the $0^{\rm th}$ order image
was determined by the {\small TGDETECT} tool, which then defined the identification
and extraction regions for the $\pm1^{\rm st}$ orders of the \meg\ and \heg.
Specifically, events within an 18\,pixel radius of the $0^{\rm th}$ order image were
assigned to $0^{\rm th}$ order, while any events that fell within $\pm 18$ pixels of the
cross dispersion direction of either the \meg\ or \heg\ spectra were assigned to that
grating arm using the {\small TG\_CREATE\_MASK} tool. In overlap regions, events
are preferentially assigned to $0^{\rm th}$, \meg, and then \heg, respectively. Within
these identification regions, events that were within $\pm1''$ of the \heg\ and \meg\
dispersion arm locations were extracted ({\small TG\_EXTRACT}) and assigned to a
given spectral order with {\small TG\_RESOLVE\_EVENTS} using the default settings,
while background spectra were extracted from regions that lay 3--8.5$''$
perpendicularly from the dispersion arm locations. The standard tools, {\small
FULLGARF} and {\small MKGRMF}, were used to create the spectral response
matrices. Finally, for a given set of gratings (\heg\ or \meg) the spectra were combined
for $\pm1^{\rm st}$ orders using \addascaspec\ for each OBSID, and where spectra
from multiple OBSIDs were combined we again used \addascaspec.

\section{Spectral Analysis}
\label{sec_spec}

Our focus on this work is on broadband spectroscopy of \ngc. All of the datasets
analysed are rebinned to have a signal-to-noise (S/N) ratio of at least 5 per energy
bin, using a combination of \specgroup\ in the \xmm\ \sas\ and custom software, and
we fit the data using \chisq\ minimisation. The \xmm\ EPIC datasets are analysed over
the 0.3--10.0\,keV range, the \chandra\ MEG and HEG gratings data over the 0.6--6.0
and 0.8--8.0\,keV ranges, respectively, and the \nustar\ FPMA/B data provide coverage
up to $\sim$25--35\,keV, depending on the dataset (above which the S/N drops below
5). We perform spectral analysis with \xspec\ v12.10.0 (\citealt{XSPEC}) throughout,
and quote parameter uncertainties at the 90\% confidence level for one interesting
parameter ($\Delta\chi^{2} = 2.7$). We allow multiplicative constants to float between
the datasets from a given epoch to account for cross-calibration uncertainties between
the different detectors, with FPMA fixed at unity as one of the detectors common to all
epochs; these constants are within $\sim$10\% of unity, as expected
(\citealt{NUSTARcal}). Throughout this work, we assume a distance of $D = 4.2$\,Mpc
to NGC\,1313 (\citealt{Mendez02, Tully16}).

\subsection{Data Organisation}
\label{sec_epochs}

The observations considered are naturally grouped into several broad epochs, largely
determined by the \nustar\ coverage; we refer to the coordinated \xmm+\nustar\
epochs as XN$i$, and the coordinated \chandra+\nustar\ epochs as CN$i$, where $i$
indicates the chronology for each combination (see Table \ref{tab_obs}; note that we
do not require strict simultaneity between the data from the relevant observatories
when defining these epochs, and that the observations combined to form each epoch
are all taken within $\sim$2 weeks of the \nustar\ exposures around which they are
defined). Figure \ref{fig_longlc} shows the \nustar\ coverage in the context of the
long-term behaviour from \ngc\ seen by the \swiftng\ (hereafter \swift; \citealt{SWIFT}),
extracted using the standard online pipeline (\citealt{Evans09}). These epochs cover
a range of fluxes, and also the period of enhanced variability seen more recently by
\swift\ (commencing MJD $\sim$ 57400).

\begin{figure}
\begin{center}
\hspace*{-0.3cm}
\rotatebox{0}{
{\includegraphics[width=235pt]{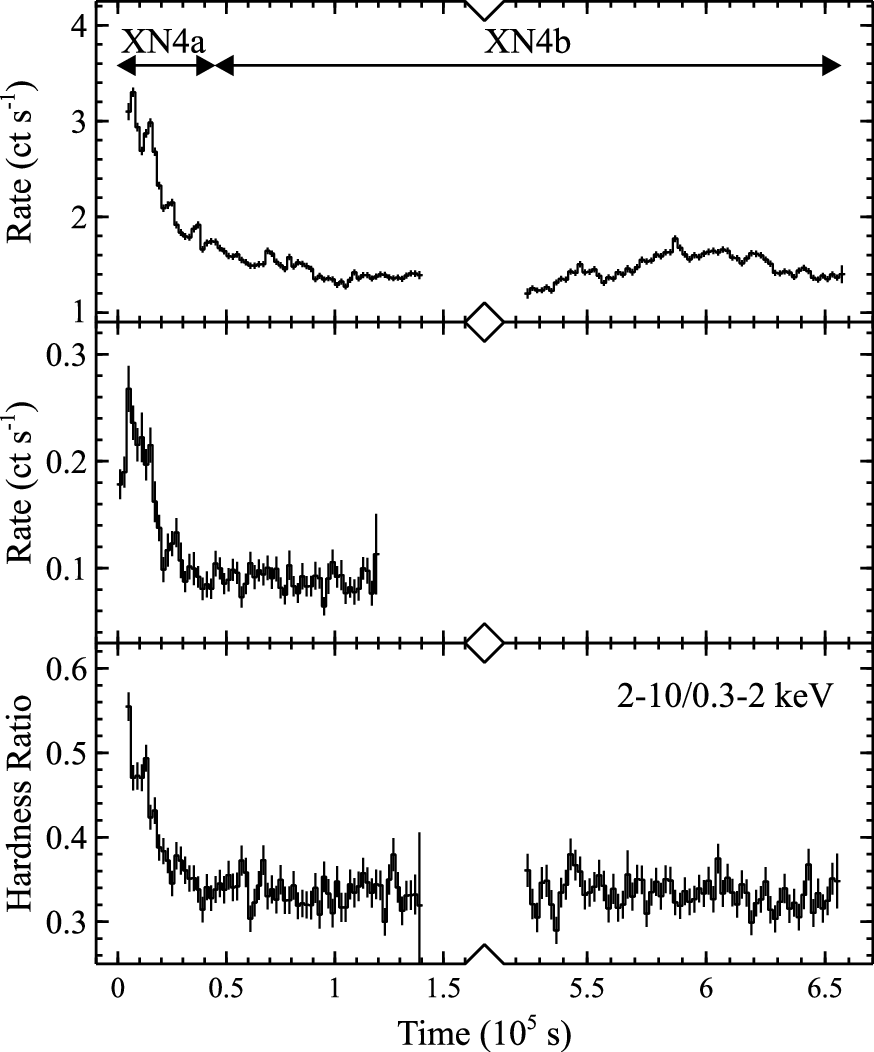}}
}
\end{center}
\vspace*{-0.3cm}
\caption{
Lightcurves from the observations that form epoch XN4 (2\,ks time bins). The top
panel shows the 0.3--10.0\,keV lightcurve observed with the \epicpn\ detector over the
two \xmm\ exposures, and the middle panel shows the 3--30\,keV lightcurve observed
with the \nustar\ FPMA detector. The bottom panel also shows a hardness ratio
computed between the 0.3--2.0 and 2.0--10.0\,keV bands, based on the \epicpn\
count rates. Clear spectral variability is seen towards the beginning of these
observations, after which the spectrum appears to stabilize (even if some flux
variability is still observed). For our spectral analysis we therefore split these data into
two sub-epochs, which we refer to as XN4a and XN4b, as indicated by the horizontal
arrows.}
\label{fig_shortlc}
\end{figure}

\begin{figure*}
\begin{center}
\hspace*{-0.3cm}
\rotatebox{0}{
{\includegraphics[width=235pt]{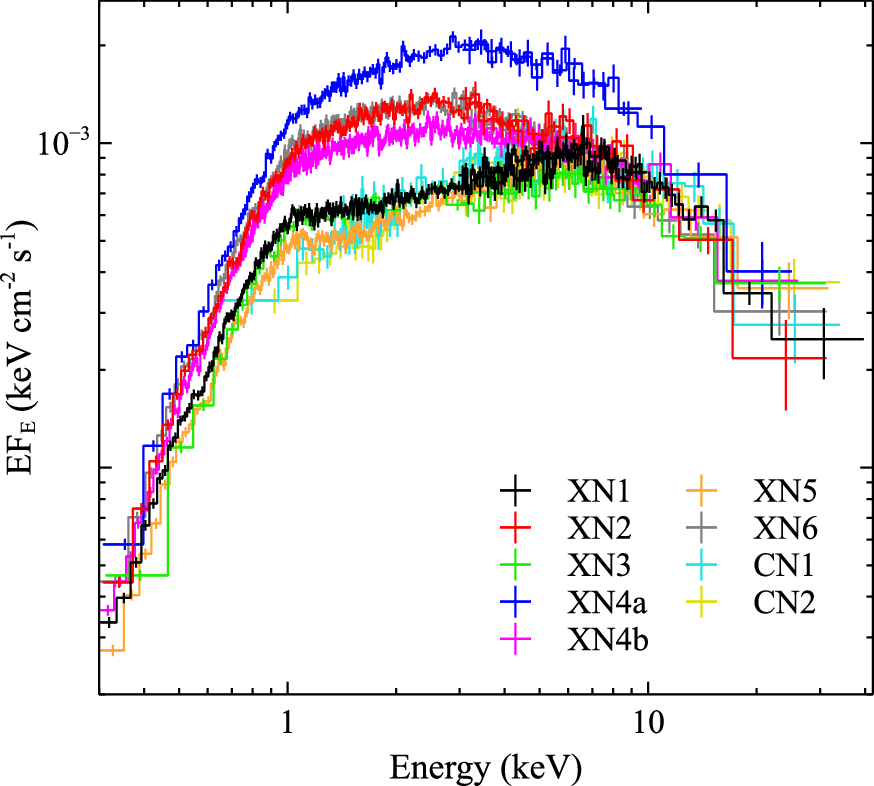}}
}
\hspace*{0.5cm}
\rotatebox{0}{
{\includegraphics[width=235pt]{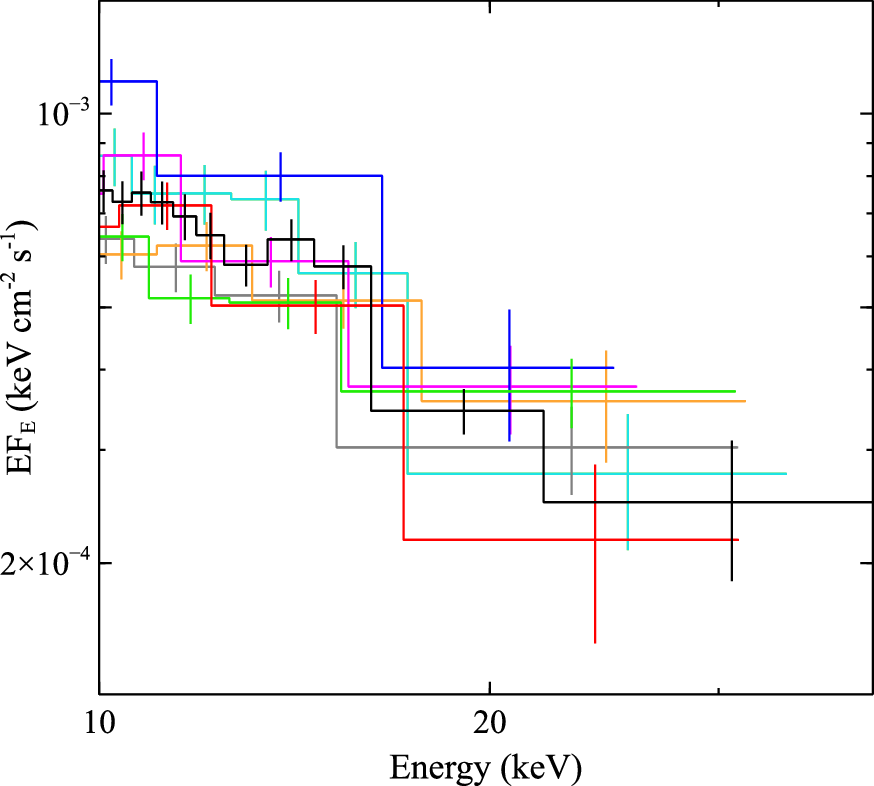}}
}
\end{center}
\vspace*{-0.3cm}
\caption{
\textit{Left:} The broadband spectral evolution displayed by \ngc. For clarity, we only
show the \epicpn\ data for \xmm\ and the MEG data for \chandra\ (where relevant),
and the FPMA data for \nustar. Given the large number of datasets shown, we show
the soft and hard X-ray coverage from each epoch in the same colour. The \xmm\
data span the $\sim$0.3--10.0\,keV range, the \chandra\ data shown span the
$\sim$0.6--6.0\,keV range, and the \nustar\ data span the $\sim$3--35\,keV range.
Despite the strong variations at lower energies (below $\sim$10\,keV), the higher
energy data (above $\sim$10\,keV) are relatively stable. \textit{Right:} A zoom in on
the high-energy ($>$10\,keV) data from \nustar\ (again, FPMA only for clarity), further
demonstrating the relative lack of variability observed at these energies in comparison
to the lower energy data. All the data have been unfolded through a simple model that
is constant with energy, and have been further rebinned for visual purposes.}
\label{fig_spec}
\end{figure*}

The two broadband observations performed in quick succession in 2012 are
considered to be a single epoch (XN1) in this work, as they both exhibit the same
spectrum and there is negligible variability within either of the observations
(\citealt{Bachetti13}). Epoch XN4 coincidentally caught the end of a relatively large
flare (see Figure \ref{fig_longlc}), with \ngc\ showing strong flux and spectral evolution
across the simultaneous \xmm+\nustar\ exposure (see Figure \ref{fig_shortlc}). We
therefore split the data from this epoch into higher and lower flux periods, which we
refer to as epochs XN4a and XN4b respectively. The XN4a spectrum is extracted from
the first 45\,ks of the simultaneous exposure (OBSIDs 0803990101 and 30302016002;
this is the point at which the high-energy ($E>3$\,keV) flux/spectral variability stabilises),
while the XN4b spectrum is extracted from the remaining data from these OBSIDs,
combined with the second \xmm\ exposure (OBSID 0803990201), which exhibited similar
flux and spectral shape. None of the other new observations considered here show
notable spectral variability within any of the individual exposures. The two \xmm\
observations taken during epoch XN5 show consistent spectra, and so are combined.
However, the two \xmm\ observations taken during epoch XN6 do show clearly different
spectra, and so in this case we only consider the \xmm\ data from the exposure taken
simultaneously with \nustar. We also note that, owing to the spread of the full \chandra\
dataset, for epochs XN5 and XN6 there are also short \chandra\ observations
contemporaneous with the \xmm\ and \nustar\ data, but in these cases we only consider
the higher S/N \xmm\ data for the soft X-ray coverage. For epochs CN1 and CN2, we
confirmed that the spectra from each of the individual \chandra\ observations grouped
together were consistent, before combining them further into the final merged spectra
used in our analysis.

\subsection{Broadband Spectral Variability}
\label{sec_spec_bb}

The nine broadband spectra extracted are shown in Figure \ref{fig_spec}. The spectral
variability exhibited by \ngc\ shows remarkable similarity to the behaviour seen from
Holmberg IX X-1 (\citealt{Walton14hoIX, Walton17hoIX, Luangtip16}), another
well-studied ULX; the variability is strong at lower energies (below $\sim$10\,keV), with
the spectrum becoming more centrally peaked at higher fluxes, but the higher energy
data (above $\sim$10\,keV) appears to remain extremely similar for all of the spectra
compiled to date. To illustrate this, we model the \nustar\ data above 10\,keV with a
simple powerlaw model. The photon indices are all consistent within their 90\%
confidence limits (fitting all of the datasets together we find $\Gamma = 3.2 \pm 0.1$),
and the 10--40\,keV fluxes only vary by at most a factor of $\sim$1.5, despite the factor
of $>$3 differences seen at $\sim$3\,keV.

As the nature of the accretor in \ngc\ remains unknown (no pulsations have been seen
from \ngc\ to date; see Appendix \ref{app_pulse}), we construct two models to
fit the broadband data that may approximate super-Eddington accretion onto a
non-magnetic accretor (either a black hole or an non-magnetic neutron star; note that
super-Eddington accretion onto a non-magnetic neutron star is expected to be
conceptually similar to the black hole case, with a standard outer disc, a funnel-like
inner region and strong winds, \eg\ \citealt{King08}, although for a given luminosity the
accretion rate relative to Eddington would naturally be more extreme in the former) and
a magnetic accretor (\ie a ULX pulsar), respectively, although we stress that these
models are still strictly phenomenological. Throughout this work, we include two neutral
absorbers in all of our modelling, the first fixed to the Galactic column of $N_{\rm{H}} =
4.13 \times 10^{20}$\,cm$^{-2}$ (\citealt{NH}), and the second free to account for
absorption local to NGC\,1313 ($z = 0.00157$). For the neutral absorption, we use the
\tbabs\ absorption model, combining the solar abundances of \cite{tbabs} and the
cross-sections of \cite{Verner96}. We also note again that \ngc\ is known to show
atomic features in both absorption and emission, particularly at low energies
(\citealt{Middleton14, Pinto16nat, Walton16ufo}). However, here we are interested in
the spectral variability of the continuum emission, and these features do not strongly
influence the broadband continuum fits, so we do not treat them here; instead the
atomic emission/absorption will be the subject of separate works (\citealt{Pinto20};
Nowak et al., \textit{in preparation}).

\begin{sidewaystable*}
  \caption[labelfont=bf]{Best-fit parameters from the nine broadband spectra currently
  available for NGC\,1313 X-1 for the models assuming a non-magnetic and a magnetic
  accretor, respectively}
  \vspace{-0.1cm}
\begin{center}
\hspace*{-0.4cm}
%\small
\begin{tabular}{c c c c c c c c c c c c}
\hline
\hline
\\[-0.2cm]
Model & \multicolumn{2}{c}{Parameter} & & & & & Epoch \\
\\[-0.3cm]
Component & & & XN1 & XN2 & XN3 & XN4a & XN4b & CN1 & XN5 & CN2 & XN6 \\
\\[-0.3cm]
\hline
\hline
\\[-0.2cm]
\multicolumn{12}{c}{Non-Magnetic Accretor Model: \tbabs $\times$ $($ \diskbb\ $+$ \diskpbb\ $\otimes$ \simpl\ $)$} \\
\\[-0.25cm]
\tbabs\ & $N_{\rm{H,int}}$\tmark[a] & [$10^{21}$ cm$^{-2}$] & $2.47^{+0.05}_{-0.08}$ & -- & -- & -- & -- & -- & -- & -- & -- \\
\\[-0.3cm]
\diskbb & $T_{\rm{in}}$ & [keV] & $0.27 \pm 0.01$ & $0.35^{+0.03}_{-0.02}$ & $0.30 \pm 0.01$ & $0.36 \pm 0.02$ & $0.33 \pm 0.01$ & $0.16^{+0.07}_{-0.05}$ & $0.25 \pm 0.01$ & $0.18^{+0.07}_{-0.06}$ & $0.32 \pm 0.01$ \\
\\[-0.3cm]
& Norm & & $17.3^{+2.0}_{-2.3}$ & $4.5^{+2.1}_{-1.6}$ & $9.9^{+1.6}_{-1.4}$ & $7.2^{+1.9}_{-2.1}$ & $6.7^{+1.2}_{-1.0}$ & $160^{+2930}_{-140}$ & $21.8^{+2.7}_{-3.1}$ & $67^{+728}_{-58}$ & $8.9^{+2.7}_{-2.2}$ \\
\\[-0.3cm]
\diskpbb\ & $T_{\rm{in}}$ & [keV] & $3.0^{+0.2}_{-0.3}$ & $1.0^{+0.3}_{-0.1}$ & $1.9^{+0.9}_{-0.5}$ & $1.6^{+0.1}_{-0.2}$ & $1.3^{+0.2}_{-0.1}$ & $3.1 \pm 0.7$ & $2.3 \pm 0.4$ & $2.9^{+0.8}_{-1.0}$ & $0.9^{+0.2}_{-0.1}$ \\
\\[-0.3cm]
& $p$ & & $0.58 \pm 0.01$ & $0.55^{+0.03}_{-0.01}$ & $0.59^{+0.04}_{-0.02}$ & $0.61^{+0.05}_{-0.03}$ & $0.55^{+0.02}_{-0.01}$ & $0.57 \pm 0.02$ & $0.62^{+0.03}_{-0.02}$ & $0.57^{+0.03}_{-0.02}$ & $0.59^{+0.05}_{-0.03}$ \\
\\[-0.3cm]
& Norm & [$10^{-3}$] & $0.8^{+0.4}_{-0.2}$ & $50^{+66}_{-30}$ & $3.1^{+4.5}_{-1.8}$ & $28^{+26}_{-7}$ & $13^{+12}_{-4}$ & $0.7^{+0.5}_{-0.4}$ & $2.6^{+3.2}_{-1.3}$ & $0.8^{+3.1}_{-0.5}$ & $96^{+165}_{-57}$ \\
\\[-0.3cm]
\simpl\ & $\Gamma$\tmark[a] & & $2.94^{+0.07}_{-0.08}$ & -- & -- & -- & -- & -- & -- & -- & -- \\
\\[-0.3cm]
& $f_{\rm{scat}}$ & [\%] & $21^{+8}_{-7}$ & $>70$ & $>32$ & $35^{+10}_{-9}$ & $>82$ & $20^{+25}_{-9}$ & $44^{+25}_{-18}$ & $36^{+44}_{-23}$ & $>70$ \\
\\[-0.3cm]
\hline
\\[-0.2cm]
\chisq/DoF & & & 12544/11559 \\
\\[-0.3cm]
\hline
\hline
\\[-0.2cm]
\multicolumn{12}{c}{Magnetic Accretor Model: \tbabs $\times$ $($ \diskbb\ $+$ \diskpbb\ $+$ \cutoffpl\ $)$} \\
\\[-0.25cm]
\tbabs\ & $N_{\rm{H,int}}$\tmark[a] & [$10^{21}$ cm$^{-2}$] & $2.53^{+0.06}_{-0.07}$ & -- & -- & -- & -- & -- & -- & -- & -- \\
\\[-0.3cm]
\diskbb & $T_{\rm{in}}$ & [keV] & $0.27 \pm 0.01$ & $0.38^{+0.04}_{-0.02}$ & $0.30 \pm 0.01$ & $0.36 \pm 0.02$ & $0.33 \pm 0.01$ & $0.17^{+0.07}_{-0.06}$ & $0.25 \pm 0.01$ & $0.18^{+0.08}_{-0.06}$ & $0.34^{+0.02}_{-0.01}$\\
\\[-0.3cm]
& Norm & & $17.5^{+2.6}_{-2.1}$ & $3.2^{+1.3}_{-1.1}$ & $10.1^{+2.0}_{-1.8}$ & $7.1^{+2.2}_{-2.0}$ & $8.1^{+1.3}_{-1.1}$ & $120^{+2580}_{-110}$ & $22.6^{+3.3}_{-2.7}$ & $61^{+1039}_{-53}$ & $6.3^{+1.7}_{-1.5}$ \\
\\[-0.3cm]
\diskpbb\ & $T_{\rm{in}}$ & [keV] & $2.9^{+0.2}_{-0.3}$ & $1.6^{+0.2}_{-0.1}$ & $2.0^{+0.4}_{-0.3}$ & $1.7 \pm 0.1$ & $1.5 \pm 0.1$ & $2.6^{+0.9}_{-0.5}$ & $2.2^{+0.4}_{-0.2}$ & $2.5^{+0.7}_{-0.5}$ &  $1.5 \pm 0.1$ \\
\\[-0.3cm]
& $p$ & & $0.56 \pm 0.01$ & $0.54^{+0.01}_{-0.02}$ & $0.57^{+0.06}_{-0.03}$ & $0.60 \pm 0.03$ & $0.55^{+0.03}_{-0.01}$ & $0.55^{+0.04}_{-0.02}$ & $0.60^{+0.03}_{-0.02}$ & $0.55^{+0.05}_{-0.02}$ & $0.56 \pm 0.03$ \\
\\[-0.3cm]
& Norm & [$10^{-3}$] & $0.7^{+0.3}_{-0.2}$ & $9.3^{+4.5}_{-3.3}$ & $2.3^{+2.8}_{-1.4}$ & $18.0^{+8.8}_{-5.8}$ & $10.8^{+5.1}_{-3.5}$ & $1.0^{+1.6}_{-0.6}$ & $2.1^{+1.2}_{-1.0}$ & $1.0^{+1.7}_{-0.6}$ & $13.8^{+5.4}_{-3.9}$ \\
\\[-0.3cm]
\cutoffpl\ & $\Gamma$ & & $0.59$\tmark[b] & -- & -- & -- & -- & -- & -- & -- & -- \\
\\[-0.3cm]
& $E_{\rm{fold}}$ & [keV] & $7.1$\tmark[b] & -- & -- & -- & -- & -- & -- & -- & -- \\
\\[-0.3cm]
& $F^{\rm{cpl}}_{2-10}$\tmark[c] & [$10^{-13}$\,\ergpcmsqps] & $5.3^{+1.0}_{-1.4}$ & $8.5^{+0.9}_{-1.1}$ & $7.9^{+0.9}_{-1.5}$ & $10.3^{+1.5}_{-1.8}$ & $11.3^{+0.8}_{-1.0}$ & $7.0^{+1.9}_{-5.3}$ & $7.2^{+1.2}_{-2.3}$ & $8.1^{+1.5}_{-3.4}$ & $8.2 \pm 0.7$ \\
\\[-0.3cm]
\hline
\\[-0.2cm]
\chisq/DoF & & & 12572/11560 \\
\\[-0.3cm]
\hline
\hline
\\[-0.2cm]
$F^{\rm{obs}}_{0.3-40.0}$\tmark[d] & \multicolumn{2}{c}{\multirow{5}{*}{[$10^{-13}$\,\ergpcmsqps]}} & $42.2 \pm 0.7$ & $59.9^{+1.9}_{-2.0}$ & $38.7^{+1.2}_{-1.3}$ & $87.2^{+3.1}_{-3.2}$ & $55.6^{+2.0}_{-1.9}$ & $41.4^{+2.0}_{-1.9}$ & $38.7^{+1.3}_{-1.4}$ & $39.1 \pm 1.7$ & $57.7 \pm 1.6$ \\
\\[-0.3cm]
$F^{\rm{obs}}_{0.3-1.0}$\tmark[d] & & & $4.1 \pm 0.1$ & $6.1 \pm 0.2$ & $3.7^{+0.2}_{-0.1}$ & $7.6 \pm 0.3$ & $5.5 \pm 0.2$ & $3.6^{+0.7}_{-1.2}$ & $3.6^{+0.1}_{-0.2}$ & $3.1^{+0.9}_{-0.7}$ & $6.1 \pm 0.2$ \\
\\[-0.3cm]
$F^{\rm{obs}}_{1.0-10.0}$\tmark[d] & & & $29.9 \pm 0.5$ & $45.3 \pm 1.5$ & $26.8^{+0.9}_{-0.8}$ & $68.5^{+2.4}_{-2.5}$ & $39.4^{+1.5}_{-1.4}$ & $29.1^{+1.0}_{-1.1}$ & $27.0 \pm 0.9$ & $26.9^{+0.9}_{-1.0}$ & $43.5 \pm 1.2$ \\
\\[-0.3cm]
$F^{\rm{obs}}_{10.0-40.0}$\tmark[d] & & & $8.1 \pm 0.4$ & $8.5^{+0.7}_{-0.8}$ & $8.1^{+0.5}_{-0.7}$ & $11.1^{+1.1}_{-1.2}$ & $10.7^{+0.7}_{-0.8}$ & $8.8^{+0.9}_{-1.1}$ & $8.1^{+0.7}_{-0.9}$ & $9.2^{+0.7}_{-1.0}$ & $8.1^{+0.5}_{-0.6}$ \\
\\[-0.3cm]
\hline
\\[-0.2cm]
$L^{\rm{int}}_{0.3-40.0}$\tmark[e] & \multicolumn{2}{c}{[$10^{39}$\,\ergps]} & $12.3^{+0.2}_{-0.3}$ & $17.6^{+0.5}_{-0.6}$ & $11.2^{+0.3}_{-0.4}$ & $24.4^{+0.8}_{-0.9}$ & $16.2^{+0.5}_{-0.6}$ & $12.3^{+2.9}_{-1.5}$ & $11.2^{+0.3}_{-0.4}$ & $11.1^{+1.9}_{-1.1}$ & $17.0^{+0.4}_{-0.5}$ \\
\\[-0.3cm]
\hline
\hline
\end{tabular}
%\vspace{0.0cm}
\label{tab_param}
\end{center}
\vspace{-0.2cm}
$^{a}$ These parameters are globally free to vary, but are linked across all epochs. \\
$^{b}$ These parameters are fixed to the average values seen from the pulsed emission
from the currently known ULX pulsars, and are common for all epochs. \\
$^{c}$ The observed flux from the \cutoffpl\ component associated with the potential
accretion column in the 2--10\,keV band. \\
$^{d}$ The total observed flux in the full 0.3--40.0\,keV band, and the 0.3--1.0, 1.0--10.0
and 10.0--40.0\,keV sub-bands, respectively (consistent for both models). \\
$^{e}$ Absorption-corrected luminosity in the full 0.3--40.0\,keV band (consistent for
both models). These values assume isotropic emission, and may therefore be upper
limits (see Section \ref{sec_dis}). \\
\end{sidewaystable*}

\begin{figure*}
\begin{center}
\hspace*{-0.3cm}
\rotatebox{0}{
{\includegraphics[width=490pt]{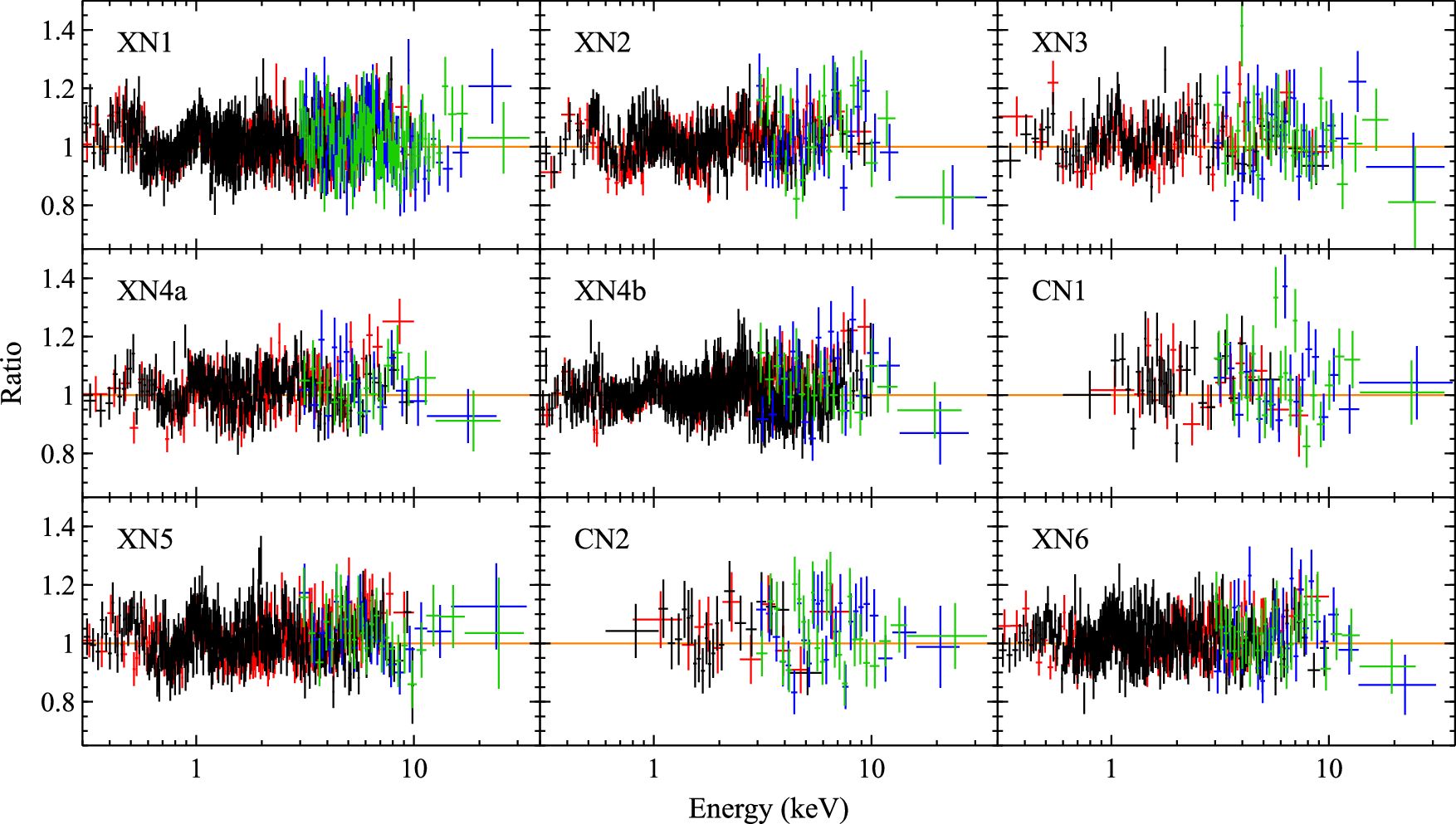}}
}
\end{center}
\vspace*{-0.4cm}
\caption{
Data/model ratios for the nine broadband spectra extracted of \ngc\ fit with the model
assuming a non-magnetic accretor. Green and blue show data from the \nustar\ FPMA
and FPMB detectors, respectively, while black and red show data from the \epicpn\ and
\epicmos\ detectors, respectively, for the XN epochs and from the MEG and HEG
gratings, respectively, for the CN epochs. As with Figure \ref{fig_spec}, the data have
been further rebinned for display purposes. The data/model ratios for the model
assuming a magnetic accretor are extremely similar to those shown here.}
\label{fig_ratioBH}
\end{figure*}

\begin{figure*}
\begin{center}
\hspace*{-0.4cm}
\rotatebox{0}{
{\includegraphics[width=490pt]{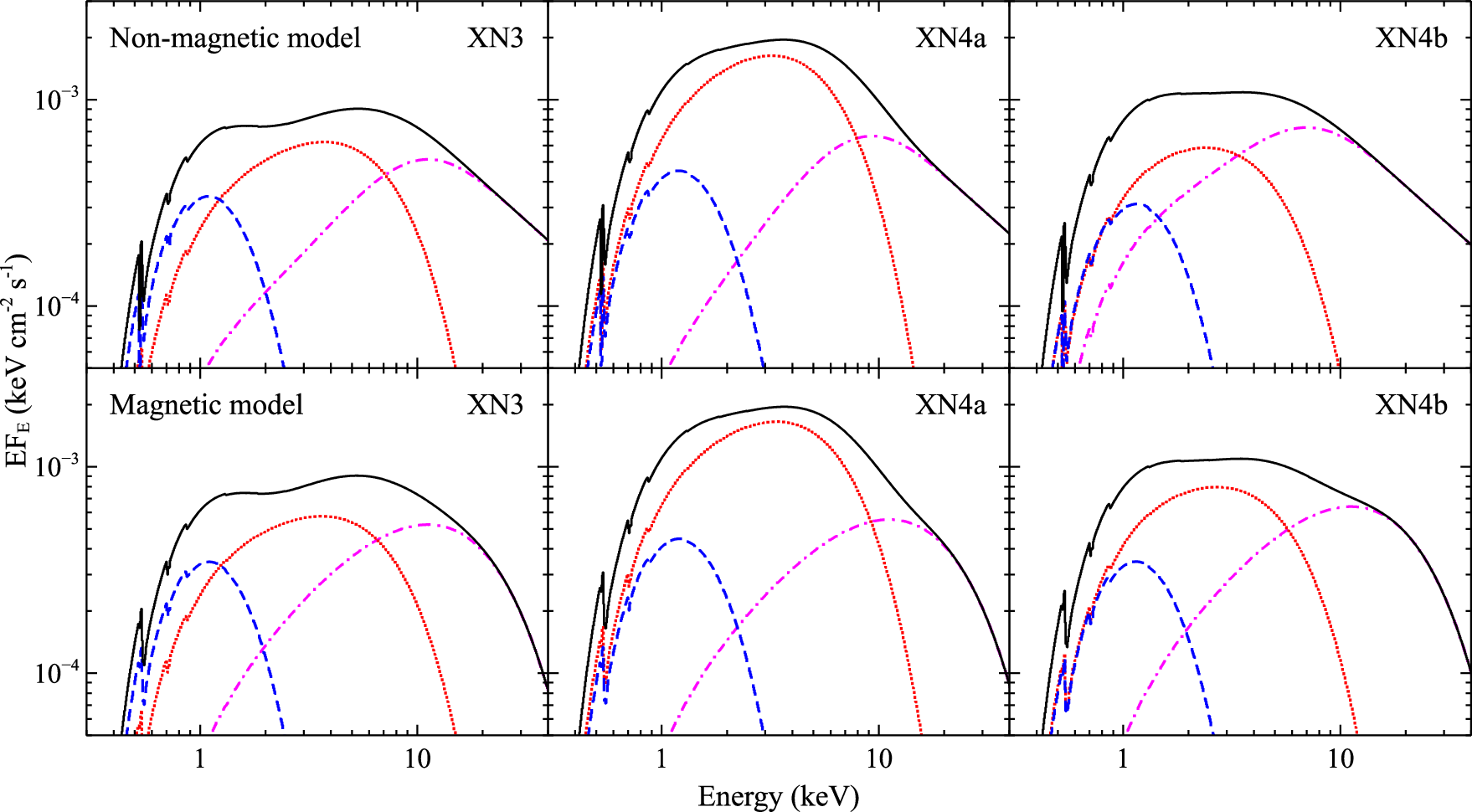}}
}
\end{center}
\vspace*{-0.3cm}
\caption{
Example model fits for the non-magnetic accretor (\textit{top panels}) and the magnetic
accretor (\textit{bottom panels}) models. We show the fits for epochs XN3 (representing
the low-flux end of the broadband spectra; \textit{left panels}), XN4a (the high-flux end
of the broadband spectra, \textit{middle panels}), and XN4b (one of the intermediate flux
spectra; \textit{right panels}). For both models, the total model is shown in solid black,
the cooler DISKBB and the hotter DISKPBB components are shown in dashed blue and
dotted red, respectively (in the black hole case, the DISKPBB component is shown as
it would appear prior to the application of the SIMPL convolution model), and the highest
energy component (either SIMPL or CUTOFFPL) is shown in dash-dot magenta. For the
non-magnetic model, the contribution from SIMPL has been calculated by subtracting
the unmodified DISKPBB component from the combined SIMPL$\otimes$DISKPBB
contribution to the fit.}
\label{fig_models}
\end{figure*}

\subsubsection{The Non-Magnetic Accretor Model}

For super-Eddington accretion onto a non-magnetic accretor, the structure of the
accretion flow is expected to deviate from the standard thin disc approximation typically
invoked for sub-Eddington accretion. As the accretion rate approaches and increases
beyond the Eddington limit, the scale-height of the inner regions of the disc is expected
to increase, supported by the increasing radiation pressure (\eg\ \citealt{Shakura73,
Abram88, Poutanen07, Dotan11}). This results in a transition from a thin disc to a
thicker flow roughly at the `spherization' radius (\rsp), the point at which the flow
reaches the Eddington limit. Radiation pressure and potentially also advection of
radiation are expected to be important for the thicker inner regions of such a flow,
which modifies the radial temperature profile -- typically parameterised as $T(r) \propto
r^{-p}$ -- of this region of the flow away from that expected for a thin disc; a standard
thin disc should have $p = 0.75$ (\citealt{Shakura73}), while a high-Eddington,
advective flow should have $p < 0.75$ (\citealt{Abram88}). Strong winds are 
also expected to be launched from the regions interior to \rsp\ (\eg\ \citealt{Ohsuga11,
Takeuchi13}), which may themselves be optically-thick (and therefore contribute
blackbody-like emission) and shroud the outer accretion flow (\eg\ \citealt{King03,
Urquhart16, Zhou19}).

We therefore use two thermal components to model the accretion flow in the scenario
that the accretor is non-magnetic, combining \diskbb\ (\citealt{diskbb}) for the outer
flow/optically-thick wind and \diskpbb\ (\citealt{diskpbb}) for the inner flow. The
\diskbb\ model assumes a thin disc profile (\ie $p = 0.75$), while the \diskpbb\ model
allows $p$ to be fit as a free parameter. This combination has frequently been applied
to explain the soft X-ray data ($E < 10$\,keV) in spectral analyses of ULXs (\eg\
\citealt{Stobbart06, Walton14hoIX, Walton17hoIX, Rana15, Mukherjee15}).

As shown in \cite{Walton18ulxBB}, even when using complex accretion disc models
such as this, all the ULXs observed by \nustar\ to date -- including \ngc\ -- require an
additional continuum component that contributes above $\sim$10\,keV. In the case of
a non-magnetic accretor, this high-energy emission would likely be associated with
Compton up-scattering of disc photons in a corona of hot electrons, as is the case in
sub-Eddington black hole X-ray binaries and active galactic nuclei (\eg\
\citealt{Haardt91}). We therefore model this emission with an additional high-energy
powerlaw tail, using the \simpl\ convolution model (\citealt{SIMPL}) to avoid incorrectly
extrapolating the powerlaw emission down to arbitrarily low energies. This component
is applied to the \diskpbb\ component, the hotter of the two components associated
with the disc, as black hole coronae are expected to be compact and centrally located
(\eg\ \citealt{Reis13corona}).

We apply this model to all of the 9 broadband datasets of \ngc\ considered in this
work simultaneously, similar to our analysis of Holmberg IX X-1
(\citealt{Walton17hoIX}). Following that work, and given the similarities between the
multi-epoch spectra at the highest and lowest energies, we assume a common
absorption column (see also \citealt{Miller13ulx}) and a common photon index for the
high energy continuum across all epochs. The global fit to the data is reasonably
good, with \chisq\ = 12544 for 11559 degrees of freedom (DoF); we give the best-fit
parameters in Table \ref{tab_param}, and show the data/model ratios for the various
datasets in Figure \ref{fig_ratioBH}. Although there are notable residuals at
$\sim$1\,keV in most cases, related to atomic emission and absorption associated
with the extreme outflow which are blended in the low-resolution spectra used here
(\citealt{Middleton14, Middleton15soft, Pinto16nat}), the shape of the continuum
emission is reasonably well reproduced. We show examples of our model fits for
epochs XN3, XN4a and XN4b (\ie covering a range of fluxes) in Figure
\ref{fig_models}.

\subsubsection{The Magnetic Accretor Model}

The model we use for the case of a magnetic accretor (\ie a ULX pulsar) is based
on that discussed in \cite{Walton18p13, Walton18ulxBB}. This consists of two thermal
blackbody components for the accretion flow outside of the magnetosphere (\rmag; the
point at which the magnetic field of the neutron star truncates the disk and the accreting
material begins to follow the field lines instead), and an exponentially cutoff powerlaw
component (\cutoffpl) for the central accretion columns that form as the material flows
down onto the magnetic poles. For the thermal components we again use the
\diskbb+\diskpbb\ combination, which has also been used in previous work on ULX
pulsars; assuming that \rmag\ $<$ \rsp, the qualitative structure of a super-Eddington
flow (thin outer disc, thick inner disc, optically-thick wind) is expected to be broadly
similar to the non-magnetic case for radii outside of \rmag. The discovery of the strong
wind in the ULX pulsar NGC\,300 ULX1 supports the conclusion that the super-Eddington
regions of the accretion flow still form in these systems (\ie \rmag \ $<$ \rsp;
\citealt{Kosec18, Mushtukov19}). For dipolar magnetic fields, this would correspond to
the lower end of the predicted range of field strengths ($B \lesssim 10^{12}$). However,
stronger fields could still be permitted with higher-order field geometries (\eg\
\citealt{Israel17, Middleton19}).

Since pulsations have not been detected from \ngc\ we can not constrain the spectral
shape of any accretion columns directly. We therefore take a similar approach to
\cite{Walton18ulxBB}, and set its spectral parameters to the average values seen from
the pulsed emission from the four ULX pulsars currently known: $\langle \Gamma
\rangle = 0.59$ and $\langle E_{\rm{cut}} \rangle = 7.9$\,keV (\citealt{Brightman16m82a,
Walton18p13, Walton18ulxBB, Walton18crsf}). These values are similar, but are not
identical to those used in \cite{Walton18ulxBB}, as NGC\,300 ULX1 had not been
discovered to be a ULX pulsar at that time; note that for this source we take the
continuum parameters from the model that includes the cyclotron resonant scattering
feature presented in \cite{Walton18crsf}. In the magnetised case, this component
provides the bulk of the high-energy ($E > 10$\,keV) emission observed by \nustar\ and
explains the high-energy excess seen even with complex disc models; the treatment of
this emission is the only difference between the non-magnetic and magnetic accretion
models used here.

As with the non-magnetic case, we apply this model to all 9 broadband datasets
considered in this work simultaneously, again assuming a common absorption column
for all epochs (the shape of the accretion column is fixed in the model, and so is also
common for all epochs). The global fit to the data is similarly good (\chisq/DoF = 
12572/11560), with the shape of the continuum similarly well described as the
non-magnetic case (we do not show the data/model ratios for the magnetic accretor
model for brevity, as they are extremely similar to Figure \ref{fig_ratioBH}), and the
best-fit parameters are again presented in Table \ref{tab_param}. We also show
examples of these model fits in Figure \ref{fig_models}; for ease of comparison, we
show the same epochs as shown for the non-magnetic accretor model.

\section{Discussion}
\label{sec_dis}

We have presented a multi-epoch spectral analysis of all of the broadband datasets
available for the bright ($L_{\rm{X}} \sim 10^{40}$\,\ergps) ULX \ngc. These datasets
combine observations taken with \xmm\ and \chandra\ in coordination with \nustar, and
span a period of $\sim$5 years. The first of these epochs, XN1, corresponds to the
data presented by \cite{Bachetti13}. From these observations we extracted 9
broadband spectra, covering the $\sim$0.5--30\,keV energy range, which probe the
spectral variability exhibited on timescales of $\sim$days to $\sim$years (see Section
\ref{sec_spec}). Several of these are broadly similar to epoch XN1 (epochs XN3, CN1,
XN5, CN2), but others probe higher fluxes and show clear differences in their spectra
(epochs XN2, XN4a,b, XN6; see Figure \ref{fig_spec}).

In a qualitative sense, the spectral variability exhibited by these observations is
remarkably similar to that seen in Holmberg IX X-1 (see Figure 1 in
\citealt{Walton17hoIX}). Strong variability is apparent at low energies (below
$\sim$10\,keV), with the spectra showing a more flat-topped profile at lower fluxes,
and becoming more centrally peaked at higher fluxes. However, at higher energies
the data pinch together and remain remarkably stable. Indeed, despite the factor of
$>$3 variations seen at $\sim$3\,keV, the 10--40\,keV fluxes only vary by a factor
of at most $\sim$1.5 (see Figure \ref{fig_spec}). A similar effect may also have
been seen at higher energies in the high-mass X-ray binary GX\,301--2, which
exhibited notable stability in the emission seen by \nustar\ above $\sim$40\,keV
despite clear variability at lower energies (\citealt{Fuerst18gx301}).

As discussed previously, ULXs are now generally expected to represent a
population of super-Eddington accretors, at least some of which are powered by
neutron stars. We therefore construct spectral models that may broadly
represent emission from a super-Eddington accretion flow. Such accretion flows are
broadly expected to be formed of a large scale-height inner funnel, a strong wind
launched by this inner funnel (which may be optically thick) and a more standard thin
outer accretion disc (which may be shrouded by the wind), so our models include two
multi-colour blackbody components with different temperatures, one for the inner
funnel and one for the outer disc/wind. These dominate the observed spectra below
$\sim$10\,keV; in general, the cooler component contributes primarily below
$\sim$1\,keV, while the hotter component contributes primarily in the $\sim$1--10\,keV
band. Similar models have frequently been applied to ULX data below 10\,keV (\eg\
\citealt{Stobbart06, Gladstone09, Walton14hoIX}). In the case of \ngc, the need for two
components below 10\,keV is visually apparent for the lower flux observations (Figure
\ref{fig_spec}). However, even for the higher flux epochs, where this is not as obvious,
the spectral decomposition found here is supported by the short-timescale variability
results presented by \cite{Kara20} for epoch XN4 (the most variable broadband epoch).
The covariance analysis presented in that work clearly shows evidence for distinct
spectral components above and below $\sim$2\,keV (see also \citealt{Middleton15}),
similar to the model used here.

However, as demonstrated by \cite{Walton18ulxBB}, when fit with such models all
ULXs observed by \nustar\ to date (including \ngc) require an additional component at
high energies to account for the \nustar\ data above $\sim$10\,keV, as the Wien tail in
accretion disc models falls off too steeply. Since pulsations have not currently been
observed from \ngc, and so the nature of the accretor in this system is not currently
known, we take two different approaches to modelling this additional high-energy
component. First, we treat it as a high-energy powerlaw tail produced by Compton
up-scattering in an X-ray corona, similar to that seen in other X-ray binary systems.
We refer to this as the non-magnetic scenario, which may be appropriate for
both black hole and non-magnetic neutron star accretors. Second, we treat it as
high-energy emission from a super-Eddington accretion column onto a magnetised
neutron star, and assume a spectral form motivated by the pulsed emission observed
from the known ULX pulsars (using the average spectral shape of their pulsed spectra
as a template). We refer to this as the magnetic scenario. 

\cite{Walton17hoIX} suggested that the broadband spectral variability seen in
Holmberg\,IX X-1, similar to that reported here for \ngc, could potentially be related to
the presence of the expected funnel-like geometry for the inner accretion flow. In such
a scenario, the funnel is expected to geometrically collimate the emission from the
innermost regions within the funnel (discussed further in Section \ref{sec_beaming}).
Regardless of the nature of the accretor (black hole or neutron star), the highest energy
emission probed by \nustar\ is usually expected to arise from these regions, either
powered by a centrally located Compton-scattering corona (\eg\ \citealt{Reis13corona}),
or a centrally located accretion column. The stability of this emission would therefore
imply that any geometrical collimation it experiences remains roughly constant, despite
the change in observed broadband X-ray flux (which would suggest a change in
accretion rate, $\dot{M}$). In principle, an increase in accretion rate would be expected
to result in an increase in the scale-height of the funnel (e.g. \citealt{King08,
Middleton15}). However, while this must happen over some range of $\dot{M}$ in order
for the disc structure to transition from the thin disc expected for standard
sub-Eddington accretion to the funnel-like geometry expected for super-Eddington
accretion, as discussed by \cite{Lasota16}, once the disc reaches the point of being
fully advection-dominated the opening angle of the disc should tend to a constant
($H/R \sim 1$, where $H$ is the scale-height of the disc at radius $R$).
\cite{Walton17hoIX} speculated that once this occurs, rather than closing the funnel
further, an increase in $\dot{M}$ instead simply increases the characteristic radius within
which geometric beaming occurs, such that emission that is already within this region
(the highest energies probed) experiences no further collimation with an increase in
$\dot{M}$, while emission from larger radii (i.e. from more intermediate energies) does
still become progressively more focused, and would exhibit stronger variability. In
essence, this idea invokes a radially-dependent beaming factor in which the beaming of
the innermost regions has saturated to explain (in only a qualitative sense) the unusual,
energy-dependent broadband spectral evolution seen from Holmberg\,IX X-I (and now
\ngc).

\subsection{Evolution of the Thermal Components}
\label{sec_bb}

\begin{figure}
\begin{center}
\hspace*{-0.3cm}
\rotatebox{0}{
{\includegraphics[width=235pt]{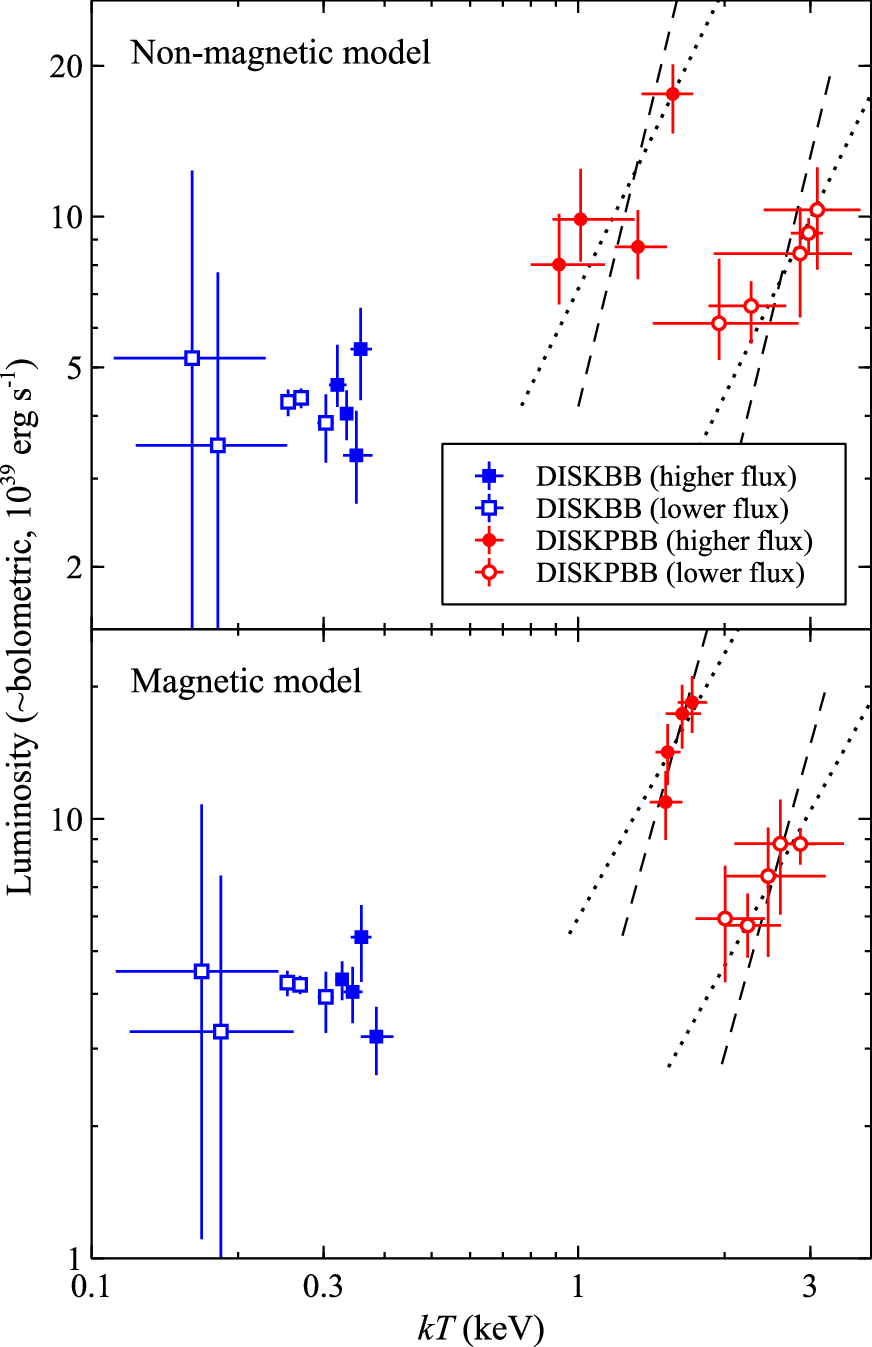}}
}
\end{center}
\vspace*{-0.3cm}
\caption{
0.001--100\,keV (\ie $\sim$bolometric) luminosity vs temperature for both the \diskbb\
(blue squares) and \diskpbb\ (red circles) components from our multi-epoch spectral
analysis with the models assuming the accretor is non-magnetic (top) and magnetic
(bottom). No clear relation is seen for the \diskbb\ component (although this may be
related to modelling issues, see Section \ref{sec_dis}). However, in both cases the
results for the \diskpbb\ component appear to show distinct tracks at high and low
luminosities (observed fluxes above and below $F^{\rm{obs}}_{0.3-40.0} = 5 \times
10^{-12}$\,\ergpcmsqps; full and open symbols, respectively), both of which are
broadly consistent with $L \propto T^4$ (indicated with the dashed lines), and also
with $L \propto T^2$ (dotted lines).}
\label{fig_LT}
\end{figure}

With this picture in mind, and with a fairly extensive, broadband dataset now available
for \ngc, we consider the behaviour of the two thermal components in each of our
models in detail. In particular, we investigate how the \diskbb\ and \diskpbb\ components
evolve in the luminosity--temperature plane. To compute the luminosity, we calculate the
intrinsic fluxes (i.e. absorption corrected, and in the case of the \diskpbb\ component for
the black hole model, corrected for the photons lost to the powerlaw tail) for each of the
thermal components individually, computed over a broad enough band to be considered
bolometric (0.001--100\,keV). The results for each component are shown in Figure
\ref{fig_LT} for both of the models considered.

We do not find any clear relationship between $L$ and $T$ for the lower temperature
\diskbb\ component. However, this is likely related, at least in part, to our treatment of
the hotter component with the \diskpbb\ model. This extends the run of temperatures
down to arbitrarily low values, which is not physically reasonable if the two thermal
components do represent radially distinct regions of the disc. This extrapolation will
naturally influence the flux of the cooler \diskbb\ component, and could certainly serve
to artificially mask any such relation, so the results presented here are not particularly
well suited for addressing this issue for the cooler component. Indeed, based on
archival \xmm\ data, we note that \cite{Miller13ulx} find evidence for a positive relation
between luminosity and temperature for the cooler \diskbb\ component when treating
the higher energy emission with a Comptonization model\footnote{Owing to the lack of
available \nustar\ data, only the 0.3--10.0\,keV energy range covered by \xmm\ was
considered by \cite{Miller13ulx}. As such, the energies dominated by the Comptonized
emission in that case correspond to the intermediate energies dominated by the
\diskpbb\ component here.}, and assuming that the seed photons come from the
\diskbb\ component (such that this emission has a low-energy cutoff at the \diskbb\
temperature). \cite{Miller13ulx} may therefore present a more accurate assessment of
the evolution of this cooler thermal component. This issue will be explored further in
future work.

The most interesting results are instead seen for the hotter \diskpbb\ component. In
both cases (the models for non-magnetic and magnetic accretors) the results cluster
into two groups, split by the observed flux. The higher-flux cases
($F^{\rm{obs}}_{0.3-40.0} > 5 \times 10^{-12}$\,\ergpcmsqps; epochs XN2, XN4a,b,
XN6) show lower temperatures on average, while the lower-flux cases
($F^{\rm{obs}}_{0.3-40.0} < 5 \times 10^{-12}$\,\ergpcmsqps; epochs XN1, XN3, XN5,
CN1, CN2) show higher temperatures. Naively, one could conclude that luminosity
and temperature are inversely correlated for this component, as also discussed for
Holmberg\,IX X-1 by \cite{Walton14hoIX}. However, when considered separately,
each of these groups appear to exhibit a positive correlation (see Figure \ref{fig_LT}).
We present Pearson's correlation coefficients ($\rho$) and null hypothesis probabilities
(i.e. the probability of no correlation; $p_{\rm{null}}$) in Table \ref{tab_corr}.
Although the data are visually compelling, the formal statistical evidence for a correlation
is not as strong for the higher luminosity/lower temperature group, in large part because
this group is only made up of 4 epochs (although we note that for the non-magnetic
accretor model, the evidence for a correlation is driven purely by epoch XN4a).
Ultimately, further observational data will be required to robustly confirm this behaviour.

\begin{table}
  \caption{Pearson's correlation coefficients and null hypothesis probabilities (no
  correlation) for the low- and high-temperature tracks for the \diskpbb\ component seen
  in Figure \ref{fig_LT} for the non-magnetic and the magnetic accretor models.}
\begin{center}
%\vspace{-0.4cm}
\begin{tabular}{c c c c c c c c}
\hline
\hline
\\[-0.25cm]
 & \multicolumn{2}{c}{Low Temp.} & \multicolumn{2}{c}{High Temp.} \\
\\[-0.35cm]
Model & $\rho$ & $p_{\rm{null}}$ & $\rho$ & $p_{\rm{null}}$ \\
\\[-0.3cm]
\hline
\\[-0.2cm]
Non-Magnetic Accretor & 0.77 & 0.23 & 0.97 & 0.005 \\
\\[-0.3cm]
Magnetic Accretor & 0.87 & 0.13 & 0.91 & 0.03 \\
\\[-0.3cm]
\hline
\hline
\\[-0.4cm]
\end{tabular}
\end{center}
\label{tab_corr}
\end{table}

Nevertheless, each of the two groups of observations are broadly consistent with
following their own distinct $L \propto T^{4}$ relationship, which would be expected for
blackbody radiation with a constant emitting area. Although the qualitative results
regarding \eg\ temperature do depend on the choice of model, and the global
uncertainties are larger in the black hole model because the shape of the high-energy
continuum is free to vary in this case, qualitatively this behaviour appears to be largely
model independent, as it is seen for both of the models considered. We have also
performed a series of other tests related to the models used, including allowing the
neutral absorption column to vary between epochs (following \citealt{Middleton15}),
replacing the cooler \diskbb\ component with a single-temperature blackbody
and linking $p$ across all of the datasets in the magnetic model, and allowing for
different photon indices for the two groups of observations in the non-magnetic model.
We still see qualitatively consistent behaviour in the hotter \diskpbb\ component to that
shown above in all of these cases. In addition, we also further tested whether the
spectral variability inferred for the \diskpbb\ component within each group of
observations is really required by the data, as this drives the two positive
luminosity--temperature correlations seen in Figure \ref{fig_LT}. To do so we re-fit the
data assuming common values for both $p$ and $T_{\rm{in}}$ for the \diskpbb\
component for the observations that make up each group, allowing only for the flux of
this component to vary within them. This significantly degrades the fit by $\Delta\chi^{2}
\sim 100$ (for 14 fewer free parameters) for both the non-magnetic and the magnetic
accretor models; F-tests imply the probabilities of these differences in fit statistic
occurring by chance are $\sim$10$^{-12}$ in both cases.

Given the concerns regarding the extrapolation of the \diskpbb\ model mentioned above,
we also investigated whether the results for this component shown in Figure \ref{fig_LT}
could be influenced by extending this component significantly outside of the observed
bandpass. Instead of using the broader 0.001--100\,keV fluxes shown in Figure
\ref{fig_LT}, we also compute the fluxes for this model component above 1\,keV, which
is primarily covered by the observed bandpass. Although the quantitative details
naturally change, the same qualitative behaviour is still seen: the observations split into
two distinct groups with their own luminosity--temperature tracks, each of which are
consistent with $L \propto T^{4}$. In the non-magnetic case the results for this higher
energy band are again not as clear-cut as the magnetic case, but the data are also still
consistent with there being two groups following separate tracks, each again consistent
with $L \propto T^{4}$. The observed behaviour therefore appears to be largely robust
to any issues regarding extrapolation of the \diskpbb\ component outside of the
observed bandpass. We also note that this demonstrates that the atomic features
associated with the wind, which are not modelled here, do not significantly influence the
observed luminosity--temperature behaviour for the \diskpbb\ component, as these
features have a very small effect on the \diskpbb\ flux above 1\,keV (at the level of a few
per cent; \citealt{Pinto20}).

%These results are shown in Figure \ref{fig_LT_1keV}.

%\begin{figure}
%\begin{center}
%\hspace*{-0.3cm}
%\rotatebox{0}{
%{\includegraphics[width=235pt]{./figs/ngc1313x1_LT_all2018_1keV.eps}}
%}
%\end{center}
%\vspace*{-0.3cm}
%\caption{
%Luminosity (above 1\,keV) vs temperature for both the \diskpbb\ (red) component
%from our multi-epoch spectral analysis with the models assuming the accretor is a
%black hole (top) and a neutron star (bottom). In both cases, despite limiting the energies
%over which the luminosity is caltulated to $>$1\,keV, similar behaviour to that seen in 
%Figure \ref{fig_LT} is again observed.}
%\label{fig_LT_1keV}
%\end{figure}

%This is in part because broadly similar values for $p$ are inferred for all epochs with both
%models, and this sets the slope of the \diskpbb\ component below its peak temperature.

Owing to the small number of observations and the relatively limited dynamic range
currently available for each track, we do not fit for formal luminosity--temperature
relations at this stage. Instead, to test for the consistency with $L \propto T^{4}$ in a
simple manner, we perform some further fits to the data in which we assume constant
inner radii ($R_{\rm{in}}$) for the \diskpbb\ component for each of the two groups of
observations. Formally we link their normalisations, which are given by $[R_{\rm{in}}/(\xi
D f_{\rm{col}}^{2})]^2 \cos(i)$, where $R_{\rm{in}}$ and $D$ are in units of km and
10\,kpc, respectively, $i$ is the inclination of the disc, $f_{\rm{col}}$ is its colour correction
factor, and $\xi$ is a further correction introduced by the inner boundary condition
assumed in the \diskbb/\diskpbb\ models ($\xi \sim 0.4$; \citealt{Kubota98, Vierdayanti08}).
The colour correction factor is a simple multiplicative correction designed to empirically
account for the complex atmospheric physics in the disc by relating its `colour' temperature
at the photosphere ($T_{\rm{col}}$) to its effective blackbody temperature ($T_{\rm{eff}}$),
and is defined as $T_{\rm{col}} = f_{\rm{col}} T_{\rm{eff}}$. Assuming that $i$ and
$f_{\rm{col}}$ are similar for the \diskpbb\ component for both groups of observations, the
ratio of their inner radii is simply related to the ratio of their normalizations, \ie
$R_{\rm{in,1}}/R_{\rm{in,2}} = \sqrt{\rm{Norm}_{1}/\rm{Norm}_{2}}$ (where subscripts 1 and
2 refer to the lower and higher temperature tracks, respectively). Although we now have
seven fewer free parameters, these fits are only worse by $\Delta\chi^{2}$ = 21 and 15 for
the non-magnetic and magnetic models, respectively, and we find that the ratios of the two
radii are $R_{\rm{in,1}}/R_{\rm{in,2}} = 4.6^{+1.2}_{-1.0}$ and $3.6^{+0.6}_{-0.7}$.

We also estimate the minimum inner radii (in the context of isotropic emission) for the
\diskpbb\ component implied by the two potential $L \propto T^{4}$ tracks by assuming
no colour correction (\ie $f_{\rm{col}} = 1$) and a face-on inclination (\ie $i = 0$). For
the non-magnetic model, these inner radii are $R_{\rm{in,1}} \sim 26$\,km and
$R_{\rm{in,2}} \sim 5.6$\,km, while for the magnetic model these inner radii are
$R_{\rm{in,1}} \sim 19$\,km and $R_{\rm{in,2}} \sim 5.2$\,km. We stress that any colour
correction and/or non-zero inclination will increase these estimates. Although the colour
correction is typically taken to be $f_{\rm{col}} = 1.7$ for sub-Eddington accretion (\eg\
\citealt{Shimura95}), for super-Eddington accretion $f_{\rm{col}} \sim 3$ may be more
appropriate (\citealt{Watari03}, \citealt{Davis19}, and in reality $f_{\rm{col}}$ may have a
radial dependence, \citealt{Soria08}). Adopting this value and an inclination of
$i = 60$\deg, for example, increases these estimates by a factor of $3^{2}/\sqrt{0.5}
\approx 12.7$ to $R_{\rm{in,1}} \sim 320$\,km and $R_{\rm{in,2}} \sim 71$\,km for the
non-magnetic model, and to $R_{\rm{in,1}} \sim 240$\,km and $R_{\rm{in,2}} \sim
66$\,km for the magnetic model. If the accretor is a 10\,\msun\ black hole, these radii
would correspond to $R_{\rm{in,1}} \sim 20$\,\rg\ (where \rg\ $= GM/c^2$ is the
gravitational radius) and $R_{\rm{in,2}} \sim 5$\,\rg, and if the accretor is a
$\sim$1.4\,\msun\ neutron star they would correspond to $R_{\rm{in,1}} \sim
115-150$\,\rg\ and $R_{\rm{in,2}} \sim 30$\,\rg. In this latter case (1.4\,\msun\ neutron
star), the radii inferred are vaguely similar to the launching radii for the two main
components of the wind inferred by \cite{Pinto20}, based on escape velocity
arguments ($\sim$50 and $\sim$300\,\rg, \ie within a factor of $\sim$2). In the former
(10\,\msun\ black hole), all of the changes inferred here would appear to be occurring
interior to the wind launching regions based on the same arguments, although the
larger radius is similarly comparable to the innermost radius for the wind (again within
a factor of $\sim$2).

It is important to note that the two groups of observations that show these different $L
\propto T^{4}$ tracks do not simply represent distinct periods of time over which these
different emitting radii remained stable. Instead, the source appears to switch
back-and-forth between them. The cadence of our broadband sampling is not
particularly constraining with regards to the transition between these two tracks; from
these data we can only determine that \ngc\ is able to move between them in the space
of a few weeks (the time between epochs XN4 and CN1). Should these two tracks
represent intrinsic evolution in the \diskpbb\ component, these two radii would therefore
imply that there are two stable geometric configurations that \ngc\ repeatedly returns
to/transitions between. It is also important to note that the epochs exhibiting higher
observed luminosities also show the larger of the two radii. This makes it unlikely the
observed behaviour is related to bulk precession of an otherwise stable (\ie constant
accretion rate) large scale-height inner flow changing our ability to view the emission
from its innermost regions (e.g. \citealt{Middleton18}). In this scenario, the smaller inner
disc radii should be observed when we can see further into the funnel, and therefore be
associated with higher observed fluxes. Furthermore, there is no hint that the long-term
variability exhibited by \ngc\ is even quasi-periodic (Figure \ref{fig_longlc}), instead
exhibiting a marked increase in seemingly aperiodic variability after MJD $\sim$ 57400
(as noted previously).

Given the presence of the two luminosity--temperature tracks, we also explore the
possibility that there are actually two distinct thermal components (each producing one
track) that are always present and, in combination, dominate the $\sim$1--10\,keV band
(such that the $\sim$0.3--40\,keV spectrum would actually be made up of four continuum
components, instead of the three used in our previous modelling). This might, for
example, represent a scenario in which there is even further radial segregation of the
accretion flow than included in our baseline models (see Section \ref{sec_rad} for further
discussion). In this picture, these two thermal components exhibit different levels of
long-term variability, such that their relative contribution changes from epoch to epoch,
and in our 3-component models the \diskpbb\ component is forced to (and has sufficient
flexibility to) primarily account for whichever of these two components dominates the
$\sim$1--10\,keV band for any particular epoch, switching its apparent properties
between the two as necessary. For brevity and simplicity, we focus on the magnetized
accretor model and replace the \diskpbb\ component with two standard \diskbb\ accretion
disc models, each of which has a normalisation linked across all the epochs considered
(\ie we assume each new \diskbb\ component varies following $L \propto T^{4}$). As such,
the full continuum model in this case consists of 3 \diskbb\ components and the \cutoffpl\
component associated with the accretion column. This actually provides a reasonable
improvement to the fit obtained with the 3-component model, with \chisq/DoF =
12454/11567 (\ie $\Delta\chi^{2} = 118$ for 7 additional free parameters); the
temperatures of the two new \diskbb\ components vary from $kT_{1} \sim 0.6-1.0$\,keV
and from $kT_{2} \sim 1.9-2.5$\,keV, respectively. In this case, the ratio of the inner radii
of the two new \diskbb\ components is $R_{\rm{in,1}}/R_{\rm{in,2}} = 6.4 \pm 0.6$, and
the minimum possible inner radii (again assuming no colour correction and a face-on
inclination) inferred are $R_{\rm{in,1}} \sim 75$\,km and $R_{\rm{in,2}} \sim 12$\,km.
We stress that these radii should be taken with a large degree of caution, as the issues
regarding extrapolation of the individual thermal models to low energies discussed
above are even further exacerbated in this case; the values are primarily presented for
completeness and reproducibility.

Although the luminosity--temperature tracks returned by our analyses are consistent
with $L \propto T^{4}$, in most cases they are also consistent with flatter
luminosity--temperature relations. In particular, most tracks are also consistent with $L
\propto T^{2}$ (also shown in Figure \ref{fig_LT}), which, even if $R_{\rm{in}}$ remains
$\sim$constant, may be expected for the inner regions of an advection-dominated disc
around a black hole (in which some of the radiated flux from these regions is trapped
by the flow and carried across the event horizon, \eg\ \citealt{Watari00}; note that this
is not possible for a neutron star accretor, as in that case the radiation must emerge in
some form). Some Galactic black hole X-ray binaries are seen to transition to a
luminosity--temperature relation similar to $L \propto T^{2}$ at high luminosities (see
\eg\ \citealt{Kubota04, Abe05}), and some ULXs also show evidence for this behaviour
(\citealt{Walton13culx}). Similar to the $L \propto T^{4}$ case, we perform additional fits
where the normalisations of the \diskpbb\ component are linked across each of the two
groups of observations in a manner that would give $L \propto T^{2}$; again with seven
fewer free parameters, we find the fits are only worse by $\Delta\chi^{2}$ = 31 and 30
for the non-magnetic and magnetic models, respectively. The global fits are therefore
marginally worse than (but still essentially comparable to) the fits assuming $L \propto
T^{4}$. Given the limited dynamic range covered by each of the tracks, the radii
estimated above assuming $L \propto T^{4}$ are likely still representative of the
characteristic emitting radii that would be inferred for each of the two groups of
observations even if $L \propto T^{2}$. However, since in this case the `true' disc
luminosity is underestimated (as some fraction is advected over the horizon), the
absolute radii would likely be underestimated (see \eg\ \citealt{Kubota04}).

In the following sections, we discuss potential physical causes for the two distinct
luminosity--temperature tracks associated with the \diskpbb\ component, and also
explore potential scenarios in which stable emitting radii could be produced in the
accretion flow for \ngc.

\subsection{Geometric Collimation and Disc/Wind Scale-Height}
\label{sec_beaming}

The above estimates for the emitting radii do not account for any geometric collimation
of the radiation that might be experienced by the emission from these thermal
components. As discussed previously, this may be expected for the inner regions of a
super-Eddington accretion flow, which, through the combination of the outer disc and
the wind, should form a funnel-like geometry. Should any of the thermal emission arise
from regions interior to this inner funnel then it should be collimated into a solid angle
set by the opening angle of this funnel, $\Omega$. By assuming no collimation, the
total luminosity emitted and in turn the emitting radii would be overestimated.
Introducing a `beaming' factor of $b = \Omega/4\pi$, such that the `observed' luminosity
inferred assuming isotropic emission, $L_{\rm{obs}}$, and the actual emitted luminosity,
$L_{\rm{int}}$, are related via $L_{\rm{int}} = bL_{\rm{obs}}$ (following \citealt{King08},
such that $b \leq 1$; we assume here that we are looking down the funnel), then should
any of the thermal emission be collimated the radii inferred from this emission would
need to be corrected by a factor $\sqrt{b}$.

Furthermore, any variations in the degree of beaming $b$ would manifest as changes
in the emitting areas/radii in our analysis. Indeed, if we consider the case where there
is more collimation of a blackbody thermal component at higher intrinsic luminosities,
as may be expected for a disk which has a larger scale-height at higher accretion
rates, we can write $b \propto L_{\rm{int}}^{-\beta}$, where $\beta > 0$ (as more
collimation corresponds to smaller $b$ in our definition). Assuming that the intrinsic
emission behaves as $L_{\rm{int}} \propto T^{\alpha}$, and that the process of
collimation does not also change $T$ (\ie $T_{\rm{int}} = T_{\rm{obs}} = T$; this will be
discussed further in Section \ref{sec_fcol}), combining this with the definition of $b$
and the scaling between $b$ and $L_{\rm{int}}$ we find that $L_{\rm{obs}} \propto
T^{\alpha(1+\beta)}$. Non negligible $\beta$ could therefore produce clear deviations
from $L_{\rm{obs}} \propto T^4$ for a constant area blackbody, with a steeper
luminosity--temperature scaling expected in this particular scenario.\footnote{Note that
this differs from the scaling discussed by \cite{King09} who, with similar assumptions
(i.e. increased beaming at higher accretion rates and no change in $T$), suggest that
increasingly beamed blackbody emission (intrinsically emitting as $L_{\rm{int}} \propto
T^4$) could result in $L_{\rm{obs}} \propto T^{-4}$. However, this essentially assumes
that the ratio $l = L_{\rm{int}}/L_{\rm{E}}$ remains constant, as in the full expression
derived $L_{\rm{obs}} \propto l^{2}/(T^{4}r^{2}b)$ (where $r$ is the emitting radius in
units of Schwarzschild radii). This ratio is clearly not constant here (as $L_{\rm{int}}$
must vary), meaning that the right-hand-side still has further temperature dependencies
that need to be accounted for (as $L_{\rm{int}} \propto T^4$). Substituting $l$ for $T$,
and again assuming that $b \propto L_{\rm{int}}^{-\beta}$, we return to the
$L_{\rm{obs}} \propto T^{4(1+\beta)}$ dependence derived here.} If both $L_{\rm{obs}}
\propto T^{4}$ and $L_{\rm{int}} \propto T^{4}$, or both  $L_{\rm{obs}} \propto T^{2}$
and  $L_{\rm{int}} \propto T^{2}$, then $b$ must be constant. Alternatively, it would still
be possible to produce $L_{\rm{obs}} \propto T^{4}$ even if $L_{\rm{int}} \propto T^{2}$
provided that $b \propto L_{\rm{int}}^{-1}$. However, in any of these cases, in order to
produce two distinct groups in the luminosity--temperature plane purely through
beaming, there would need to be a sharp transition in $b$ at some point. This is
naturally problematic for any model invoking progressive changes in the opening angle
of a large scale-height inner flow. In addition to being problematic for models invoking
progressive changes in the scale-height of the disc, it is similarly unclear how the picture
of a progressively changing radial beaming profile suggested by \cite{Walton17hoIX} for
Holmberg\,IX X-1 would be able to explain the two distinct groups of observations seen
for \ngc.

A model invoking a larger scale-height at higher accretion rates could instead
potentially produce a sharp transition in the luminosity--temperature plane should our
viewing angle be close to the opening angle of the flow at lower luminosities, such that
by increasing the scale-height of the disc/wind the innermost regions of the flow are 
suddenly obscured by the regions at larger radii, resulting in a larger inner radius being
inferred at higher luminosities.\footnote{Here we assume that the inner regions are fully
obscured, such that none of the emission from these regions is visible to us.}
For this to be a plausible explanation for the broadband behaviour, the obscuring regions
would need to correspond to those contributing the lower temperature parts of the
\diskpbb\ component that are still visible in the higher flux observations. A toy model for
the transition in the luminosity--temperature plane in this scenario is shown in Figure
\ref{fig_toymodels} (\textit{left}). However, if the thermal emission from the inner regions
of the disc is suddenly obscured, one would naturally expect a central corona/accretion
column to be similarly obscured, but the high-energy \nustar\ data are rather stable.

\subsection{Scattering Losses in a Wind}
\label{sec_wind}

\begin{figure}
\begin{center}
\hspace*{-0.3cm}
\rotatebox{0}{
{\includegraphics[width=235pt]{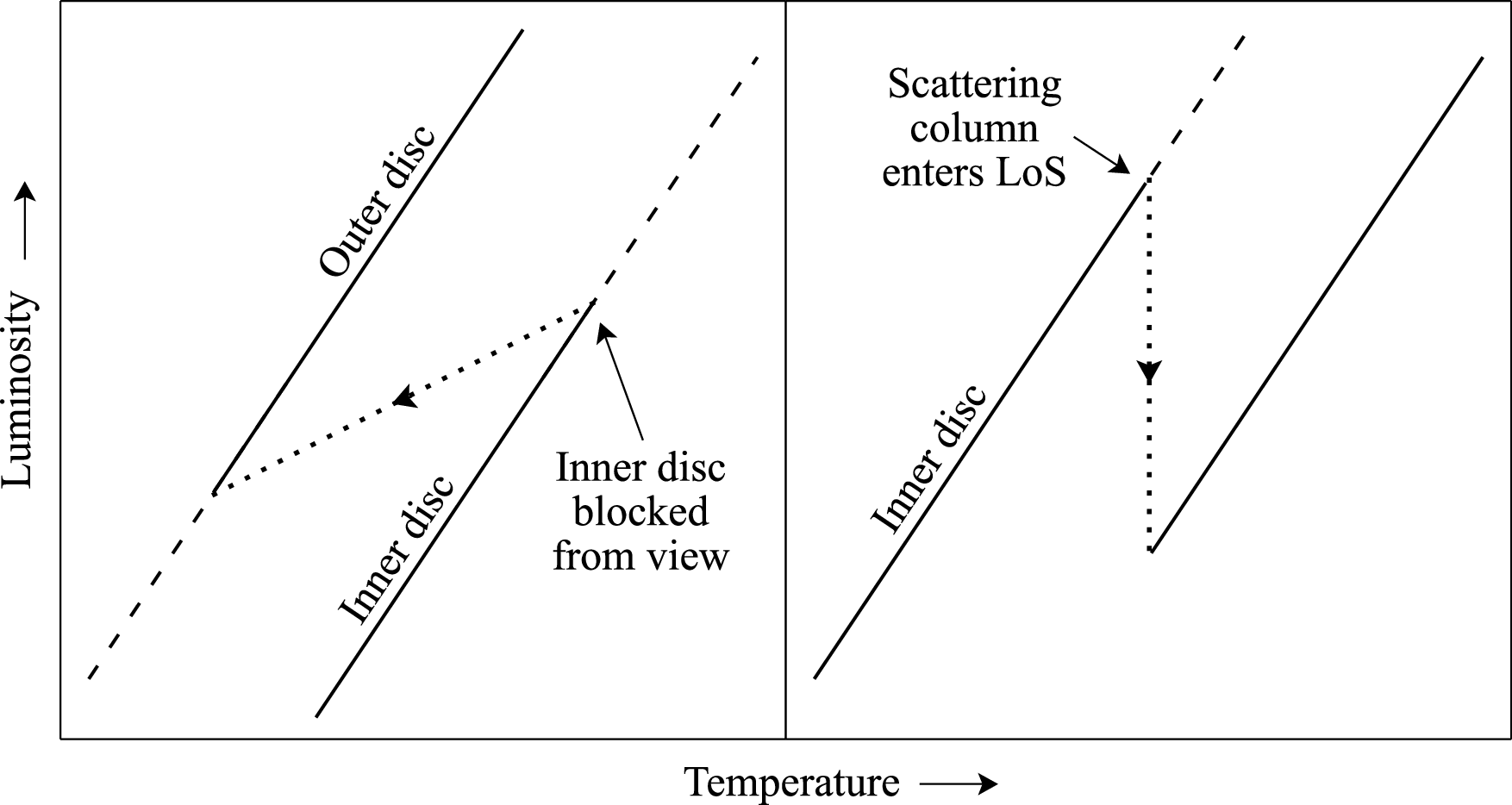}}
}
\end{center}
\vspace*{-0.3cm}
\caption{
Toy models for the luminosity--temperature plane for the scenarios in which the inner
disc is suddenly fully obscured by the outer disc/wind (\textit{left panel}; discussed in
Section \ref{sec_beaming}) and in which the emission from the inner disc is scattered
by a highly ionised wind which still permits some of the emission to escape
(\textit{right panel}; discussed in Section \ref{sec_wind}).}
\label{fig_toymodels}
\end{figure}

Alternatively, it may be the case that the high-temperature track is actually a smooth
continuation of the low-temperature track, but that above some observed luminosity we
view the emission through an ionised disc wind, which results in an apparent decrease
in the observed flux due to losses associated with electron scattering.\footnote{This is
conceptually similar to the possibility of the cooler outer disc blocking the hotter inner
regions discussed in Section \ref{sec_beaming}, but here some emission from the inner
regions is still able to be transmitted to the observer.} A toy model for the transition in the
luminosity--temperature plane in this scenario is also shown in Figure \ref{fig_toymodels}
(\textit{right}). Indeed, \ngc\ is now known to launch a powerful disc wind
(\citealt{Middleton15soft, Pinto16nat, Pinto20}), which at least at times has a
highly-ionised component along our line-of-sight (\citealt{Walton16ufo}). To test this
scenario we perform some further fits in which we assume a single normalisation for the
\diskpbb\ component for all epochs, and apply a \cabs\ component to this emission for
the observations that make up the higher-temperature track, again focusing on the
magnetic accretor model for brevity. \cabs\ accounts for flux losses due to electron
scattering, and is characterised by an effective column density for the scattering medium,
which we link across the high-temperature observations in order to preserve a common
intrinsic $L_{\rm{obs}} \propto T^4$ scaling for both the high- and low-temperature tracks.
We find that this provides an equivalently good fit to the model with two linked \diskpbb\
normalisations (see Section \ref{sec_bb}), with \chisq/DoF = 12604/11567, and that the
scattering column density required to bring about the drop in flux inferred from an
extrapolation of the low-temperature track to the observed high-temperature track is
$N_{\rm{H}} = 3.1^{+0.3}_{-0.4} \times 10^{24}$\,\pcmsq.

The scattering column required is significantly larger than the best-fit column densities
inferred for any of the components of the wind in \ngc\ reported to date. For the
moderately ionised components that contribute the features detected at $\sim$1\,keV
by the RGS, $N_{\rm{H}} \sim 10^{21-22}$\,\pcmsq\ (\citealt{Pinto16nat, Pinto20})
\footnote{Formally \cite{Pinto16nat} also find that a highly-ionised absorber with a very
large column density ($N_{\rm{H}} \sim 10^{24}$\,\pcmsq) provides a significant
improvement to the fit for the archival \xmm\ data when allowed to have a large blueshift
as well as very large velocity broadening. However, the continuum model used in that
work did not properly incorporate the high-energy curvature known to be present in ULX
spectra (\eg\ \citealt{Gladstone09, Walton18ulxBB}), and robustly confirmed by \nustar\
for \ngc\ (\citealt{Bachetti13}); instead of modelling discrete atomic features, this ionised
absorption component primarily served to introduce the required high-energy curvature
into the continuum at the highest energies probed by \xmm. Indeed, re-analysis of these
data with a more suitable continuum model finds no evidence for an absorption
component with these properties (\citealt{Pinto20}).} For the highly ionised component
that provides the iron K absorption seen in archival \xmm\ and \nustar\ data (which are
primarily made up of epoch XN1), $N_{\rm{H}} \sim 10^{23}$\,\pcmsq\
(\citealt{Walton16ufo}), although it is worth noting that there is a local minimum in the
parameter space for the highly ionised component that does extend up to column
densities comparable to that inferred above ($N_{\rm{H}} \sim 10^{24}$\,\pcmsq).
Fixing the \cabs\ column density to $10^{23}$\,\pcmsq\ in this scenario significantly
degrades the fit ($\Delta\chi^{2}$ = 165 for one fewer free parameter, giving a negligible
F-test probability of a chance improvement), as the scattering losses are very small; the
\cabs\ component is therefore unable to reproduce the required drop in flux, and the
model instead tries to produce this drop by introducing large differences in $p$ for the
two groups, resulting in a notably worse fit.

As noted above, this scenario would require there to be a sharp transition in the wind
properties along our line-of-sight to explain the observed behaviour. Furthermore, if both
tracks follow $L_{\rm{obs}} \propto T^4$, or both follow $L_{\rm{obs}} \propto T^2$, then
the wind properties would need to be bi-modal, such that we are either viewing the
central regions through a negligible scattering column, or through basically the same
column whenever our line-of-sight intercepts the wind. Although \cite{Middleton15soft}
do find that the residuals at $\sim$1\,keV imprinted by the wind are weaker in archival
\xmm\ observations that would lie on the higher-flux, lower temperature track, the
evolution appears to be a continuous function of the observed hardness of the source.
Indeed, \cite{Pinto20} find that, although there are some differences in the properties of
the wind between observations in the two tracks, both still show broadly similar
absorption from the moderately-ionised components in the RGS data. In order for this
scenario to be plausible, the local minimum at $N_{\rm{H}} \sim 10^{24}$\,\pcmsq\
reported in \cite{Walton16ufo} would likely need to be the correct solution, and this
highly ionised component would also need to show much stronger variability between
the two tracks than these more moderately ionised components. The work so far on the
wind properties in the new 2017 campaign has focused on the RGS band, and is not
particularly sensitive to the highly ionised component that would be most relevant for
this scenario (the current absorption analysis does not exclude this scenario,
\citealt{Pinto20}); this will be further addressed in future work. Alternatively, it could be
that the scattering medium is a fully ionised component of the wind, such that it
does not imprint any discernible absorption features. However, fully ionising a column of
$N_{\rm{H}} \sim 3 \times 10^{24}$\,\pcmsq\ such that there is no significant absorption
opacity below 10\,keV is obviously challenging, particularly if the \diskbb\ temperature
($\sim$0.2--0.4\,keV) represents the characteristic temperature of the wind. Regardless,
as with the scenario in which the outer disc fully blocks the inner disc, it is similarly
difficult to explain the lack of strong long-term variability at the highest energies in this
scenario under the assumption that this emission comes from the most compact regions.

\subsection{The Colour Correction Factor and Down-Scattering}
\label{sec_fcol}

Beyond geometric considerations, which all have difficulty explaining the stability of the
highest energy emission, it may also be possible to produce two apparently distinct
luminosity--temperature tracks by varying the colour correction factor, $f_{\rm{col}}$,
introduced by the atmosphere of the disc (previously we assumed a single value for all
of the available data). However, similar to the above cases, in order to do so
$f_{\rm{col}}$ would need to have two distinct values that it varies between, or at the
very least exhibit a sharp jump at some point in its evolution with accretion rate. This
would result in a sudden change in the observed temperature at a given luminosity.
However, none of the works that have tried to consider how this should vary with
accretion rate have shown an obvious indication for such a sudden jump
(\citealt{Shimura95, Davis19}). Furthermore, the general expectation among these
works is that $f_{\rm{col}}$ should increase with increasing accretion rate, which should
result in the highest temperatures being observed at the highest luminosities. Again,
this is not the case, so it is not clear that this is a realistic possibility either.

In addition to the atmospheric corrections associated with the disc, if the emission
from the inner regions is geometrically collimated by the outer disc/wind, the cooler
temperatures associated with these regions could result in significant down-scattering
(\eg\ \citealt{Middleton15}). This would lower the temperature observed even if the bulk
of the scattering occurs away from our line-of-sight, such that $T_{\rm{col}} = f_{\rm{ds}}
T_{\rm{obs}}$, where $f_{\rm{ds}} \geq 1$ (we parameterise this separately as
$f_{\rm{ds}}$ to make the distinction with $f_{\rm{col}}$). As the wind launching radius
should increase with increasing accretion rate (see Section \ref{sec_rad}), we might
expect a scenario in which there the down-scattering introduces a larger effect at higher
luminosities, such that $f_{\rm{ds}} \propto L^{\gamma}$ with $\gamma > 0$. For a
given luminosity--temperature relationship prior to any down-scattering of the form $L
\propto T_{\rm{col}}^{\alpha}$, and neglecting the effects of beaming here, we find that
increasingly strong down-scattering at higher luminosities should modify the observed
luminosity--temperature relation to $L \propto T_{\rm{obs}}^{\alpha/(1-\alpha\gamma)}$.
This kind of down-scattering relation would therefore either steepen the observed
luminosity--temperature relation if the trend remains positive (since this requires $\alpha
\gamma < 1$), or reverse the trend to give a \textit{negative} luminosity--temperature
relation. This may therefore provide another means by which it would be possible to
have observed $L_{\rm{obs}} \propto T_{\rm{obs}}^{4}$ even if $L_{\rm{int}} \propto
T_{\rm{int}}^2$. However, as with all of the other scenarios considered, producing the
two groups of observations would require a sharp jump in the degree to which
down-scattering influences the observed emission, and our naive expectation is that
this should vary smoothly with accretion rate. Furthermore, reprocessing of the emission
from the inner regions by the outer disc/wind is generally considered to be related to the
lowest temperature emission (\ie the \diskbb\ component) given the apparent
connection between ULXs and ultraluminous supersoft sources (ULSs; e.g.
\citealt{Urquhart16, Pinto17}).

\subsection{Super-Eddington X-ray Binaries: Key Radii}
\label{sec_rad}

Having considered a variety of different possible mechanisms by which the two
luminosity--temperature tracks could be produced by either geometric or atmospheric
corrections to the inner disc emission, none of which are particularly compelling, we
now consider the possibility that we are seeing further distinct key radii/regions of the
accretion flow even within the energy range covered by the \diskpbb\ component in our
models. For super-Eddington accretion onto either a black hole or a neutron star there
are several key radii to consider: the inner edge of the disc, the spherization radius (\rsp;
the point at which the disc transitions from a standard thin outer disc to the thick inner
disc expected for super-Eddington accretion, which should also correspond to the point
at which strong, radiatively-driven winds start to be launched), and the photon trapping
radius (\rtrap, the radius interior to which photons are primarily advected inwards rather
than released locally; note that \rtrap\ $<$ \rsp, \citealt{Poutanen07}).

\subsubsection{Black Hole Accretor}

For a black hole accretor, the accretion disc is expected to extend in to the innermost
stable circular orbit (\risco) at high accretion rates. This is set purely by the mass and
spin of the black hole, neither of which can be changing significantly over the course of
the observations considered here, so the inner regions of the accretion flow would be
expected to have a constant inner radius even if the accretion rate onto the black hole
changes. \risco\ would therefore be a natural candidate for the smaller of the two
potential stable radii in \ngc\ ($R_{\rm{in,2}}$; we note again that, even if $L \propto
T^{2}$ for these observations, this would imply a stable inner radius if the disc is in the
advection-dominated regime). Indeed, the rough estimates for the absolute value of this
radius are appropriate for the innermost stable circular orbit of a typical 10\,\msun\ black
hole, depending on its spin.

However, in contrast to \risco, based on standard accretion theory the spherization
radius is expected to scale with the accretion rate, \ie \rsp\ $\propto$ $\dot{m}$ (where
$\dot{m}$ is the accretion rate in units of the Eddington accretion rate;
\citealt{Shakura73, Poutanen07}). The trapping radius is also expected to have a
similar dependence, \ie \rtrap\ $\propto$ $\dot{m}$ (\citealt{Lasota16rev}). Should the
lower-temperature track exhibit a constant radius, and be related to either of these radii
in the flow, this would imply that the accretion rate is constant. However, changes in
\mdot\ are almost certainly required to produce the flux and temperature variations
observed. Allowing instead for a varying emitting radius that increases with luminosity, 
as would be expected for both \rsp\ and \rtrap, we can write $R \propto L^{\delta}$
(where $\delta > 0$), and show that in this case $L \propto T^{4/(1-2\delta)}$ (assuming
$L \propto R^{2}T^{4}$, since we are considering the regions outside \rtrap). This scenario
should either again steepen the observed luminosity--temperature relation away from $L
\propto T^{4}$ (for $0 < \delta < 0.5$) or even reverse the trend such that $L_{\rm{obs}}$
and $T$ are anti-correlated (for $\delta > 0.5$). Given that positive correlations are
observed, we may therefore expect the lower temperature track to have a steeper
luminosity--temperature relation than the higher temperature track if this is associated
with either \rsp\ or \rtrap, particularly if the inner regions vary as $L \propto T^{2}$.
Although the data do not obviously require this, given the limited observations we cannot
exclude this possibility (and there is maybe a weak hint that this is the case for the results
from the magnetic model). However, if the effects of e.g. beaming are more pronounced
for the higher-temperature track, then this difference could be reduced.

Assuming that the \diskbb\ component comes from the disc/wind at the transition to the
super-Eddington flow at \rsp, one interesting possibility is therefore that we are seeing a
further transition in the disc at \rtrap\ manifested in the behaviour of the \diskpbb\
component. The decreased variability at high energies could then be a result of higher
accretion rates leading to both stronger winds and increased photon trapping, such that
mass accretion rate fluctuations in the outer disc lead to a much weaker response than
would naively be expected from the inner regions, as discussed in \cite{Middleton15}.
This could also be qualitatively similar to the high-energy `saturation' effects discussed
by \cite{Feng19}.

\subsubsection{Non-Magnetic Neutron Star Accretor}

For a neutron star accretor, the inner edge of the accretion flow is either set by the
surface of the neutron star for non- or weakly-magnetised neutron stars (modulo the
presence of any boundary layer; the innermost stable circular orbit is likely similar to or
smaller than the radius expected for a typical neutron star: \risco\ $\sim$ 12\,km, while
\rns\ $\sim$ 13\,km; \citealt{Miller19, Riley19}) or the magnetospheric radius for
strongly-magnetised neutron stars (\rmag; the point at which the magnetic field of the
neutron star truncates the disc and the in-falling material is forced to follow the field
lines instead). Similar to \risco, the radius of the neutron star cannot be evolving
significantly over the course of our observations, so this may also be a plausible
explanation for one of the potentially stable radii if the neutron star is non-magnetic.
While the value of the inner radius estimated for the smaller of the two potential stable
radii ($R_{\rm{in,2}}$) is larger than the characteristic radius expected for a neutron
star, it may still be possible for the potential geometric beaming effects discussed above
to bring the two into consistency (although this would require $0.05 \lesssim b \lesssim
1$, depending on $f_{\rm{col}}$ and $i$, and the lower end of this range would represent
moderately extreme beaming).

However, unlike the black hole case, radiation cannot be advected over the horizon
here; while advection may still be a relevant process within some regions of the disc,
this radiation must emerge somewhere, presumably in the boundary layer where the
disc meets the neutron star surface. If this boundary layer behaves as an extension of
the disc, it is not clear that it would be possible for such a scenario to produce an $L
\propto T^{2}$ scaling; even with advection $L \propto T^{4}$ may be expected in this
case. Alternatively, though, this boundary layer may be the origin of the highest energy
emission observed by \nustar\ (\ie $E > 10$\,keV), as the material shocks at the neutron
star surface. If this is the case, it may still be possible for an advection-dominated disc
around a neutron star to produce $L \propto T^{2}$, and we may again expect the lower
temperature track to have a steeper luminosity--temperature relation. However, in this
case the lack of long-term high-energy variability would likely imply that the winds are
removing the majority of any accretion rate fluctuations before they reach these
regions (\citealt{Middleton15}).

\subsubsection{Magnetic Neutron Star Accretor}

In the classic picture of accretion onto a magnetized neutron star, \rin\ = \rmag, and
\rmag\ is set by the combination of the magnetic field of the neutron star and the
accretion rate: \rmag\ $\propto B^{4/7} \dot{M}^{-2/7}$ (\citealt{Lamb73, Cui97}).
However, this assumes that the disc is gas-pressure dominated, which is not expected
to be the case for super-Eddington accretion. \cite{Chashkina19} construct a model for
accretion onto a magnetised neutron star that extends to high accretion rates, building
on their previous model that accounts for radiation pressure (\citealt{Chashkina17}) by
further incorporating the effects of advection and outflows following the prescription of
\cite{Poutanen07}. Within this framework, at low accretion rates the disc is dominated
by gas pressure, then as the accretion rate increases the disc becomes dominated by
radiation pressure, and as the accretion rate increases further, the disc eventually
becomes dominated by the effects of advection. At low accretion rates,
\cite{Chashkina19} find that the evolution of \rmag\ with accretion rate does follow
something close to the model of \cite{Lamb73}. Interestingly, though, while the disc is in
the radiation-pressure dominated regime \cite{Chashkina19} find that \rmag\ actually
becomes constant with varying accretion rate, before exhibiting a weak dependence
again when the disc becomes advection-dominated (although in this latter case the
dependence is weaker than that seen in the gas-pressure regime). During the
radiation-pressure dominated regime, instead of pushing \rmag\ further in, the increase
in accretion rate instead primarily causes the scale-height of the disc outside of \rmag\
to increase, and the dependence of \rmag\ on accretion rate re-appears in the
advection-dominated regime because the local Eddington limit then prevents the
scale-height of the disc increasing beyond $H/R \sim 1$ (see also \citealt{Lasota16}).

Emission from a radiation-pressure dominated accretion disc around a magnetized
neutron star may therefore offer another promising explanation if the innermost radius
of the disc is stable. In this scenario, we would have \rmag\ $>$ \rtrap, and so we would
expect the disc to vary intrinsically as $L \propto T^{4}$. By itself, this would not explain
the two luminosity--temperature tracks, but \cite{Mushtukov15} suggest that the
accretion curtains that link the disc outside \rmag\ to the central accretion columns may
be optically-thick for the super-Eddington accretion seen in the known ULX pulsars, and
would thus emit multi-colour blackbody spectra, which could also potentially provide the
hotter \diskpbb\ track in this scenario. However, in this case the scale-height of the inner
disc would be variable, resulting in variable collimation for the emission arising from
regions interior to \rmag. In this case, we may therefore expect the hotter
luminosity--temperature track to exhibit a steeper scaling, but if anything the opposite
currently appears more likely.

Alternatively, if the disc were advection-dominated in this scenario (\rtrap\ $>$ \rmag),
we would now be in a situation where the inner radius would decrease with increasing
luminosity (as opposed to the non-magnetic cases where it remained constant). This
would correspond to $\delta < 0$, following our earlier notation. Therefore, it may be
that a luminosity--temperature relation even flatter than $L \propto T^{2}$ would be
expected in this case. Assuming the lower temperature track relates to regions outside
of \rtrap, the expectation that these data would show a steeper luminosity--temperature
relationship would therefore be even stronger. Again, we note that there is a weak
indication this may be the case with the current data and the magnetic accretor model.

Any radiation advected through the disc in this scenario would also escape from the
inner walls of the truncated, large scale-height flow at \rmag. If the surface and
mid-plane temperatures of the disc differ significantly (as may be the case if
$f_{\rm{col}} \sim 3$), and a significant fraction of the intrinsic flux is advected, then
this could appear as a distinct component in the observed spectrum at lower
temperatures than the innermost emission from the surface. It may even be possible
that this emission is the cause for the two luminosity--temperature tracks, particularly
if it is strongly enhanced via beaming. However, this would require that the advected
emission through the disc be more variable than the emission from the upper surface.
In turn, this would suggest that strong accretion rate fluctuations are surviving through
to \rmag, such that similarly strong variability may again be expected for the accretion
column, contrary to what is observed. Nevertheless, even if this emission does not
dominate the lower-temperature observations, it may make a non-negligible
contribution. Assuming instead that the lower temperature track primarily represents
emission from larger radii, as long as changes in the wind can efficiently prevent
changes in the accretion rate from reaching the innermost regions
(\citealt{Middleton15}), it may still be possible to explain the reduced high-energy
variability.

Given the arguments presented above, should both tracks imply stable radii, it is
also tempting to consider a scenario in which the inner radius of a radiation-pressure
dominated disc transitions between the magnetosphere and the surface of the neutron
star. Indeed, recent simulations of accretion onto magnetised neutron stars show two
distinct regimes for actively accreting neutron stars (\citealt{Parfrey17a, Parfrey17b}).
One is the `standard' regime in which the magnetic field truncates the accretion flow at
\rmag, forcing the in-falling material along the field lines, while at even higher accretion
rates the accreting material fully crushes the magnetosphere, and the disc extends in
to the surface of the neutron star. If a neutron star accretor can sharply transition
between these two regimes it may be possible to produce two well-defined groups of
observations that each follow separate luminosity--temperature tracks. However, in any
scenario along these lines we would expect to see the smaller of the two stable inner
radii to be associated with the higher luminosity observations, which as noted
previously is not the case. Furthermore, the two states simulated by \cite{Parfrey17b}
likely represent two snapshots of the gradual evolution of the magnetospheric radius
with accretion rate described by \cite{Chashkina19},\footnote{It is worth noting that the
simulations presented by \cite{Parfrey17b} do not include radiation, and so cannot
formally reproduce the radiation-pressure dominated regime described by
\cite{Chashkina19}. Initial efforts to include radiation in such have been undertaken by
\cite{Takahashi17}, but currently only a single accretion rate has been presented.} so it
is not clear that a sharp transition between these states should be expected in any
case. Lastly, the lack of strong variability is likely also problematic for this scenario, as
the central accretion columns producing the high-energy emission would not likely be
present if the magnetosphere were fully crushed.

\section{Summary and Conclusions}

We have presented results from the major coordinated X-ray observing program on the
ULX \ngc\ performed in 2017, combining \xmm, \chandra\ and \nustar, focusing on the
evolution of the broadband ($\sim$0.3--30.0\,keV) continuum emission. Clear spectral
variability is observed, but this is markedly suppressed above $\sim$10--15\,keV. This is
qualitatively similar to the broadband spectral evolution seen in Holmberg IX X-1.
Furthermore, when fit with accretion disc models designed to represent super-Eddington
accretion, the various observations trace out two distinct tracks in the
luminosity--temperature plane. Larger emitting radii and lower temperatures are seen at
higher observed fluxes. However, each of these tracks individually show positive
correlations between $L$ and $T$, and are consistent with an $L \propto T^{4}$ scaling,
as would be expected for blackbody emission with a constant emitting area, and also
with an $L \propto T^{2}$ scaling, as may be expected for an advection-dominated disc
around a black hole accretor. The limited dynamic range covered for each track currently
prevents us from distinguishing between these possibilities; further broadband
observations spanning a broader range in flux are required to confirm the precise nature
of these luminosity--temperature relations.

We have considered a variety of different possible scenarios that may be relevant for
super-Eddington accretion onto \ngc\ in order to try and explain this unusual behaviour,
allowing for both a neutron star and a black hole accretor (since the nature of the
system is not known at the current time). These include geometric changes (precession
of the flow, beaming of the radiation, obscuration of the inner regions), as well as
atmospheric effects (colour correction in the disc atmosphere, down-scattering in the
wind). However, based on our current understanding, the majority of these are expected
to vary smoothly with accretion rate, making it challenging to produce the sharp
transition required to explain the two luminosity--temperature tracks, and many would
predict that higher temperatures should be seen at higher luminosities, in contrast to the
observations. One of the more promising scenarios among this set with regards to the
luminosity--temperature behaviour is that, as the accretion rate increases, the
scale-height of the outer disc/wind also increases (as expected for super-Eddington
accretion) and blocks some of the inner (and hottest) regions from view. However, this
is difficult to reconcile with the reduced level of variability at the highest energies, under
the typical assumption that this emission arises from the most compact regions (either
in a central Compton scattering corona or an accretion column). Should this be the
cause of the luminosity--temperature behaviour, we may need to invoke other origins
for the highest energy emission, e.g. bulk-motion Comptonization in the more extended,
diverging outflow.

Alternatively, it may be that we are seeing evidence for even further radial stratification
of the flow than included in our simple 2-component model for the thermal emission
from the disc/wind. One interesting possibility here is that we are seeing evidence for a
further transition in the disc at the photon trapping radius, in addition to the spherization
radius. This could plausibly explain both the two distinct luminosity--temperature tracks,
and through a combination of outflows and advection, explain the suppressed variability
seen at the highest energies.

%
%Nature has found some way to produce this unholy mess, but I can't easily tell you how.
%
%In brief, I don't think we can have the variability of the hotter thermal component(s) be
%dominated by variable obscuration (either through full obscuration or by scattering) AND
%also have the highest energy emission come from the most compact regions. One or
%more of these assumption probably has to give. Maybe the high-energy emission here
%comes from bulk motion Comptonisation in a diverging wind?
%

\section*{ACKNOWLEDGEMENTS}

The authors would like to thank the reviewer for their positive and thorough feedback,
which helped to improve the final version of this manuscript.
DJW and MJM acknowledge support from an STFC Ernest Rutherford Fellowship.
CP and FF acknowledge support from ESA Research Fellowships.
HPE acknowledges support under NASA contract NNG08FD60C. TPR was funded
as part of the STFC consolidated grant ST/K000861/1.
NW acknowledges CNES support for this work.
CRC and MN acknowledge support from the Smithsonian Astrophysical Observatory
(SAO) contract SV3-73016 to MIT, which is in turn supported by NASA under contract
NAS8-03060 for the Chandra X-ray Center. This research has made use of data
obtained with \nustar, a project led by Caltech, funded by NASA and managed by
NASA/JPL, and has utilized the \nustardas\ software  package, jointly developed by
the ASDC (Italy) and Caltech (USA). This research has also made use of data
obtained with \xmm, an ESA science mission with instruments and contributions
directly funded by ESA Member States.

%{\it Facilities:} \facility{NuSTAR}, \facility{XMM-Newton}, \facility{Chandra}

\appendix

\section{Pulsation Searches}
\label{app_pulse}

We have undertaken basic pulsation searches for all of the 2017 observations of \ngc\
taken with high time resolution instruments (\ie \xmm\ and \nustar). To do so, all times
were transferred to the solar barycenter using the DE200 solar ephemeris. For the
\xmm\ observations, we then scanned all the available EPIC-pn light-curves (including
both the 2017 campaign and archival datasets) for pulsations by following the procedure
implemented in \cite{Sathyaprakash19}. We did not find any significant detections above
the 3$\sigma$ confidence level, and all the candidate signals below this threshold were
found to have substantially different frequencies (above $\sim$ 1\,Hz). 

We also ran a timing analysis to search for pulsations in the data from all \nustar\
OBSIDs. First, we estimated the power density spectrum in all these observations.
Each of the observations were divided into 512-s segments, each of which was defined
to be fully covered by the \nustar\ good time intervals in order to avoid spurious
frequencies arising from the aliasing of orbital occultation gaps. We then averaged the
PDS calculated in each segment and looked for signals exceeding the false-alarm
probability of white noise, taking into account the number of trials (\ie the number of
spectral points). This was done using \hendrics\ (based on \stingray;
\citealt{STINGRAY}), but did not return any significant signals. We also ran an
accelerated search for pulsations using the \presto\ software package
(\citealt{PRESTOaccel}). After creating light curves with a binning time of 0.1 ms, we
produced binary files compatible with \presto. We ran the {\small ACCELSEARCH} tool,
with and without the search over a second derivative (a ``Jerk’’ search;
\citealt{PRESTOjerk}). All candidates from the accelerated search were analysed with
the {\small PREPFOLD} tool and the diagnostic plots were inspected by eye. The
majority of the low-significance candidate pulsations are beats of the sampling frequency 
and an instrumental 890\,Hz oscillation (related to \nustar\ housekeeping operations).
They are distributed on a continuum, powerlaw-like spectrum in the sigma vs spin period
plot, so that they form a `noise' level to which we can compare credible candidates. The
higher-significance candidates (exceeding the previously described noise level) have
very low periods, none of which are consistent between observations or even between
different segments of the same dataset. These are all found to be aliases of the orbital
occultation data gaps (there is currently no way in \presto\ to correct for these missing
data).

\bibliographystyle{/Users/dwalton/papers/mnras}

\bibliography{/Users/dwalton/papers/references}

\label{lastpage}

\end{document}